\documentclass[11pt,a4paper]{article}
\usepackage{amsmath,amssymb}
\usepackage{epsfig,graphicx}
\usepackage{subfigure}
\usepackage{graphicx}
\usepackage{rotating}
\usepackage{bm}
\usepackage{axodraw}
\usepackage{color}
\usepackage[sort&compress,square]{natbib}

\newcommand{\be}{\begin{equation}}
\newcommand{\ee}{\end{equation}}
\def\bea{\begin{eqnarray}}
\def\eea{\end{eqnarray}}

\def\Tr{{\rm Tr}}
\def\tr{{\rm tr}}
\def\sst{\scriptscriptstyle}
\def\Nf{{N_{f}}}

\def\uno{\mbox{1 \kern-.59em {\rm l}}}

\let\vev=\Vev

\begin{document}
\date{\mbox{ }}
\title{{\normalsize  IPPP/08/91 DCPT/08/182\hfill\mbox{}\hfill\mbox{}}\\
\vspace{2.5 cm} \Large{\textbf{On the Diversity of Gauge Mediation:\\
Footprints of Dynamical SUSY Breaking}}}
\vspace{2.5 cm}
\author{Steven Abel, Joerg Jaeckel, Valentin V. Khoze and Luis Matos\\[3ex]
\small{\em Institute for Particle Physics Phenomenology, Durham University, Durham DH1 3LE, UK}\\[1.5ex]
\small{\em }s.a.abel, joerg.jaeckel, valya.khoze, l.f.p.t.matos@durham.ac.uk\\[1.5ex] }
\date{}
\maketitle

\vspace{2ex}
\begin{abstract}
\noindent Recent progress in realising dynamical supersymmetry
breaking allows the construction of simple and calculable
models of gauge mediation. We discuss the phenomenology of the
particularly minimal case in which the mediation is direct, and show
that there are generic new and striking predictions. These include
new particles with masses comparable to those of the Standard Model
superpartners, associated with the pseudo-Goldstone modes of the
dynamical SUSY breaking sector. Consequently there is an unavoidable
departure from the MSSM. In addition the gaugino masses are
typically significantly lighter than the sfermions, and their mass
ratios
can be different
from the pattern dictated by the gauge couplings in standard (i.e. explicit) gauge mediation. We investigate
these features in two distinct realisations of the dynamical SUSY breaking sector.
\end{abstract}

\newpage

\section{Introduction} \label{sec:intro}

In the run-up to
the LHC,
the implementation of supersymmetry (SUSY) breaking and its mediation are coming
under renewed scrutiny. Attention has recently focussed on how
dynamical supersymmetry breaking (DSB) can be achieved,
and how its effects can subsequently be transmitted to the Standard Model
sector. This interest was stimulated by the observation of
Intriligator, Seiberg and Shih (ISS) \cite{ISS}
that DSB readily occurs in very simple and calculable
SQCD-like models.

Clearly it is the interaction of the DSB sector
with the visible sector that plays a crucial role in BSM phenomenology.
However, constructing a viable model that incorporates both sectors
presents a twofold problem: both
SUSY breaking {\em and} $R$-symmetry breaking need to be transmitted
to the visible sector. The $R$-symmetry plays an important role because
supersymmetry breaking requires unbroken $R$-symmetry
(in a generic theory) \cite{Nelson:1993nf},
which is at odds with the fact that (Majorana) gauginos must
have a mass that violates $R$-symmetry.

In principle the metastable models of ISS
can circumvent this theorem by allowing
a moderate and controlled amount of $R$-symmetry breaking.
What ISS reminded us is that, because the Nelson-Seiberg
theorem \cite{Nelson:1993nf} applies only to the global vacuum of the theory,
we are at liberty to generate gaugino masses if we are prepared to
tolerate a certain amount of metastability.

New avenues for gauge mediation were consequently opened up. One
phenomenological application came shortly after with
Ref.~\cite{Murayama:2006yf}, which noted that because the ISS
model breaks supersymmetry in a magnetic Seiberg-dual formulation,
the couplings of {\em explicit} messenger fields to the DSB sector
is naturally suppressed by powers of $\Lambda_{\sst ISS}/M_{Pl}$
where $\Lambda_{\sst ISS}$ is of order the Landau pole in the
theory\footnote{Strictly speaking it is the mass scale governing
the identification of the composite meson $Q\tilde{Q}$ of the
electric ISS theory with the elementary meson $\Phi$ of the
magnetic theory, $Q\tilde{Q}=\Lambda_{\sst ISS} \Phi$ }. Thus the
magnetic theory can maintain an approximate $R$-symmetry even if
the underlying electric theory has no $R$-symmetry and is generic.
The phenomenology of this scenario is similar to standard gauge
mediation although, because of the weakness of the coupling to the
DSB sector, the scale of supersymmetry breaking has to be much
higher than is normally assumed. An alternative method of dealing
with the $R$-symmetry question is to assume that it is broken
spontaneously. Several examples of both one-loop and tree-level
$R$-symmetry breaking were developed in
Refs.~\cite{Shih:2007av,Abel:2007jx,Cho:2007yn,Shihextraordinary,Abel:2007nr,Carpenter:2008wi,Sun:2008va}
and very minimal models of {\em direct} mediation (i.e. where the
``quarks'' of the dynamical SUSY breaking sector play the role of
messengers)~\cite{Affleck:1984xz,Murayama:1997pb,ArkaniHamed:1997jv,Poppitz:1996fw,Luty:1998vr}
based on  a "baryon"-deformation of the ISS model were developed
in Refs.~\cite{Abel:2007jx}. These followed earlier developments
in
Refs.~\cite{Amariti:2006vk,Dine:2006gm,ACJK,Craig:2006kx,Fischler:2006xh,heat4,Dine:2006xt,Csaki:2006wi,Aharony:2006my,Kitano:2006xg,Ferretti:2007ec,Brummer:2007ns,Haba}.

A distinction between the phenomenology of the two kinds of model was drawn in
Ref.\cite{Abel:2007nr} where it was noted that, whereas the explicit mediation
models are rather similar to standard gauge mediation, the direct mediation
models can differ significantly, with much heavier scalar superpartners
than usual. (Benchmark points were presented in Ref.~\cite{Abel:2007nr}  to
support this, and also to show that a baryon-deformed ISS model coupled to the
MSSM model, provides a fully calculable system of broken supersymmetry.)
Several questions remain however which we will address in this paper.
At first sight, one might suspect that this kind of spectrum indicates a
residual approximate $R$-symmetry in the model, possibly because it is
broken spontaneously at one-loop -- indeed this would seem to be a mildly split
version of the argument presented in {Ref.~\cite{ArkaniHamed:2004yi}.  On closer inspection however,
the precise reason for the suppression of gaugino masses is a little more complicated. Moreover the ISS-like
DSB sector itself may become phenomenologically important
because, in direct mediation, it contains states charged under
SM gauge groups that are light (typically of order 1 TeV).

This paper follows the story to its logical conclusion: we will
catalogue the possible ways that such supersymmetry and
$R$-symmetry breaking ends up in the visible sector, using various
exemplary models of different types of breaking and gauge
mediation (direct or indirect). We conclude that direct mediation
generically yields phenomenology quite different from normal gauge
mediation. This is due partly to the $R$-symmetry and partly to
the fact that in direct mediation one of the fields to which the
messengers couple is a pseudo-Goldstone mode. Generally the
visible sector phenomenology ranges from a mildly split spectrum
to a very heavy scalar (split-SUSY like) spectrum. In addition, in
direct mediation the pseudo-Goldstone modes are expected to enter
the visible spectrum, giving a rich source of new TeV mass
particles associated with the SUSY breaking sector. This is
similar to the effects of light pseudomoduli which have been found
in \cite{Oz} in the context of explicit R-symmetry breaking
models.

We will also note that
explicit mediation and spontaneously broken $R$-symmetry can be problematic
in ISS-like models, due to the possibility that messengers become tachyonic.
Thus the
best prospect for indirect gauge mediation (i.e. with explicit messengers)
is explicit $R$-symmetry breaking
of the form discussed in Ref.~\cite{Murayama:2006yf}.

 \subsection{Overview}

Our point of reference for the present paper is, the model of Ref.~\cite{Abel:2007jx}, which introduced
into the ISS superpotential a so-called "baryon deformation" that projected out some of the
$R$-symmetry to satisfy the condition that some fields get $R$-charges different from 0 and 2
\cite{Shih:2007av}. This baryon-deformed, or \emph{ISSb} model,
is a natural deformation of the
ISS model which at tree-level has  a runaway to {\em broken} supersymmetry. Upon adding the Coleman-Weinberg
contributions  to the potential, the runaway direction is stabilized at large field values where the
$R$-symmetry is spontaneously broken. If part of the flavour symmetry of the ISS model is gauged and identified
with the parent $SU(5)$ of the Supersymmetric Standard Model (SSM), the
magnetic quarks can then be enlisted to play the role of messengers, providing an extremely simple model of direct
mediation. Moreover it was shown in Ref.~\cite{Abel:2008tx} that the Landau pole problem that usually plagues
direct gauge mediation can be avoided: this is because the ISS model itself runs into a Landau pole above which a
well-understood electric dual theory takes over. This results in a nett reduction in the effective number of messenger
flavours coupling to the SSM {\em above} the scale $\Lambda_{\sst ISS}$, and this in turn prevents the
Standard Model coupling running to strong coupling -- a scenario dubbed "deflected gauge unification".

In this paper we would like to generalize these observations
to a much wider class of models.
In order to do this we will begin in the following
section by introducing an alternative way to  break the $R$-symmetry
of the ISS model spontaneously, by adding
a meson term (with some singlet fields) to the superpotential.
We call this the "meson-deformed" ISS
model, or \emph{ISSm} model.
This bears some resemblance to the class of models
considered previously in Ref.~\cite{Carpenter:2008wi}, although now
the $R$-symmetry is broken radiatively rather than at tree-level,
thus allowing it to be somewhat simpler. We will show how
supersymmetry and $R$-symmetry are broken, using both an analytic
tree-level analysis and then a numerical minimization of the full
Coleman-Weinberg potential.

We then, in Section~\ref{sec:meson-med}, go on to show how the supersymmetry
breaking can subsequently
be mediated, first in Subsection~\ref{sec:meson-ex}  with an explicit
(indirect) mediation where we introduce
an additional messenger sector, and then in Subsection~\ref{sec:meson-dir} with direct mediation. In the
former case the phenomenology is similar to the standard gauge mediation
picture \cite{Giudice:1998bp} --
that is gauginos and scalars have similar masses governed by a
single scale and related by functions
of the gauge couplings and group theory indices. In particular
the absence of tachyonic messenger states requires the explicit mediation
model to lie in this regime, and we argue that this is likely to
require additional {\em explicit} $R$-symmetry
violating messenger mass terms.
(In this case the spontaneous $R$-symmetry breaking that we have so carefully arranged would
become irrelevant.) Thus
indirect gauge mediation in the ISS model works best with explicit
$R$-symmetry breaking of the form discussed in Ref.~\cite{Murayama:2006yf}.

On the other hand in Subsection~\ref{sec:meson-dir} we
find that the directly mediated meson-deformed
model {\em does} avoid tachyons without explicit $R$-symmetry breaking and
gives phenomenology of a different sort, similar to that of
the baryon-deformed model: the gaugino masses are suppressed.
We then turn to one of the main goals of the paper which is
to answer the question of {\em why} gauginos are so light
compared to the scalar spectrum, and
to see if this is a generic feature of spontaneously broken $R$-symmetry, or is more to do with how the
mediation occurs. In fact we shall see that both aspects play a role: it occurs only with direct
mediation, but is also related in a rather
indirect way to the fact that the $R$-symmetry is broken spontaneously. The $R$-symmetry and the equations of motion
enforce certain relations between the $F$-terms which makes the
gaugino masses cancel at leading order in messenger mass-insertions.
Once the one-loop Coleman-Weinberg contributions to the potential are included
the $F$-terms violate these classical relations and generate non-trivial contributions
to gaugino masses at the leading order in mass-insertions (the
numerical analysis shows that the precise behaviour is rather complicated). It is
this effect which gives the leading contribution to the gaugino masses.

In addition, in this gaugino suppressed regime, we shall find that the contribution from the
adjoint pseudo-Goldstone modes ,
whose mass is lifted only at one-loop, can become important. In Subsection~\ref{sec:meson-dir-a}
we consider this second question in more detail. We shall see that the pseudo-Goldstone modes
can have a significant impact on the SSM mass spectrum, and indeed their mediated contribution to the gaugino
mass can be dominant in precisely the direct mediation models where the
gauginos are light.
This is because their one-loop suppressed mass makes them behave like a mediating sector with a
correspondingly lower messenger scale. Moreover in order to give TeV scale
SUSY breaking in the visible sector, the scale of hidden SUSY breaking
is typically taken to be order $\frac{16\pi^2}{g^2}$~TeV or slightly higher.
Thus the one-loop suppressed masses of the
pseudo-Goldstone modes are typically around the scale of
SUSY-breaking in the visible sector.
This is a generic prediction: models of direct gauge mediation predict additional
(with respect to the MSSM)
scalar and fermion states
in the visible sector, corresponding to pseudo-Goldstone modes,
whose masses are close to the weak scale. We also note that the gaugino masses
do not necessarily obey the usual relation where their mass ratios scale with the
ratios of the coupling constants.
Finally in Section~\ref{sec:baryon-d} we repeat the entire analysis for the baryon-deformed model. We find that
the picture is similar.

Thus we conclude that indirect explicit mediation gives
the standard picture of gauge mediation and that explicit $R$-symmetry
breaking masses for messengers are most likely required.
On the other hand direct mediation
leads to a mass spectrum with heavy scalars and suppressed gaugino masses.
Here the $R$-symmetry breaking {\em{can}} be spontaneous, in which case
the pseudo-Goldstone modes can play a significant role in the mediation and
in the visible sector phenomenology.

\section{Meson-deformed ISS theory as the susy-breaking sector} \label{sec:meson-d}

As summarized in the Introduction,
there are two simple types of deformation one might contemplate adding to
the ISS model in order to make it spontaneously break $R$ symmetry
and generate Majorana gaugino masses in the visible sector. The first was presented in Refs.~\cite{Abel:2007jx,Abel:2007nr}
and corresponds to adding a baryonic operator to the original model.
That possibility will be examined and extended in Section \ref{sec:baryon-d}. Here
we will discuss an alternative possibility which is to add appropriate mesonic
deformations to the original model.

We will work entirely in the low-energy magnetic (i.e. relevant to collider phenomenology)
description of the ISS model \cite{ISS};
it contains $N_{f}$ flavours of quarks and
anti-quarks, $\varphi$ and $\tilde{\varphi}$ respectively, charged
under an $SU(N)$ gauge group, as well as an $N_{f}\times N_{f}$
meson $\Phi_{ij}$ which is a singlet under this gauge group.
This is an $SU(N)$ gauge theory with $N=N_f-N_c$ which is weakly coupled in the IR.
The ISS superpotential is given by
\begin{equation}
W_{\sst ISS}=h(\Phi_{ij}\varphi_{i}.\tilde{\varphi_{j}}-\mu_{ij}^{2}\Phi_{ji})\, .
\label{ISSsuperp}
\end{equation}
The coupling $h$ is related to the different dynamical scales in
the electric and magnetic theories (or equivalently the mapping
between the two gauge couplings). The parameter $\mu_{ij}^{2}$ is
derived from a Dirac mass term $m_{Q} Q \tilde{Q}$ for the quarks
of the electric theory: $\mu^{2}\sim\Lambda_{\sst ISS} m_{Q}$
where the meson field $\Phi_{ij}= \frac{1}{\Lambda_{\sst ISS}} Q_i
\tilde{Q_j}$ and where $\Lambda_{\sst ISS}$ is the Landau pole of
the theory. Equation \eqref{ISSsuperp} gives the tree-level
superpotential of the magnetic ISS SQCD theory; there is also the
non-perturbatively generated \be W_{\rm dyn}\, =\, N\left(
\frac{\det_{\sst \Nf} h \Phi}{\Lambda_{\sst
ISS}^{\Nf-3N}}\right)^\frac{1}{N}\, , \label{Wdyn} \ee which gives
negligible\footnote{The only exception to this is the $R$-axion
field. For this the explicit $R$-symmetry breaking contained in
$W_{\rm dyn}$ gives a contribution to the mass~\cite{Abel:2007jx}
which importantly facilitates the evasion of astrophysical
bounds~\cite{Kim:1986ax,Raffelt:2006cw,Raffelt:1996}. For a recent
discussion of the $R$-axion detection prospects at the LHC see
\cite{Ibe}.} contributions to physics around the SUSY-breaking
vacuum.

The flavour symmetry of the magnetic model is initially $SU(N_{f})$.
When we do \emph{direct} mediation, see Section \ref{sec:meson-dir}, an $SU(5)_f$ subgroup of this symmetry
is gauged and identified with the parent $SU(5)$ of the Standard
Model, so that $N_{f} \ge N+5$. On the other hand \emph{indirect} mediation, considered in Section \ref{sec:meson-ex},
involves the introduction of \emph{explicit } messengers and in that
case $N_{f}$ is a free parameter.

To visualise the the general set-up, let us first consider a
simple example, which is appropriate for either case:
we shall choose
an $SU(2)$ gauge group for the magnetic dual theory and $N_{f}=7$ flavours,
with the flavour symmetry broken by $\mu_{ij}$ to $SU(2)_{f}\times SU(5)_{f}.$ We will refer to this
as the 2-5 model which was the also the prototype model\footnote{We will show momentarily that
the meson-deformed ISS model actually requires a slightly more general flavour-breaking pattern which can be
described by 1-1-5 and  2-2-3 models or their generalisations.
For baryon-deformations all of these models, including the simplest 2-5 scenario will also work.}
considered in Refs.~\cite{Abel:2007jx,Abel:2007nr}.
The matter field decomposition under the $SU(2)_{f} \times SU(5)_{f}$ flavour subgroup and the charge assignments
under
$SU(2)_{gauge} \times SU(2)_{f}\times SU(5)_{f}\times U(1)_B \times U(1)_R$ are given
in Table~\ref{fieldstableM}.
\begin{table}
\begin{center}
\begin{tabular}{|c|c|c|c|c|c|}
\hline
2-5 Model
&{\small $SU(2)_{\rm mg}$} &
{\small $SU(2)_f$}&
$SU(5)_f$& {\small $U(1)_{B}$} &
{\small $U(1)_{R}$}\tabularnewline
\hline
\hline
$\Phi_{ij}\equiv\left(\begin{array}{cc}
Y & Z\\
\tilde{Z} & X\end{array}\right)$&
{\bf 1}&
$\left(\begin{array}{cc}
Adj +{\bf 1} & \bar\square\\
\square & {\bf 1}\end{array}\right)$&
$\left(\begin{array}{cc}
{\bf 1} & \square\\
\bar{\square} & Adj+{\bf 1}\end{array}\right)$& 0 &
2
\tabularnewline
\hline
{\small $\varphi\equiv\left(\begin{array}{c}
\phi\\
\rho\end{array}\right)$}&
$\square$&
$\left(\begin{array}{c}
\bar{\square}\\ {\bf 1} \end{array}\right)$&
$\left(\begin{array}{c}
{\bf 1}\\
\bar{\square}\end{array}\right)$& $\frac{1}{2}$ &
$R$
\tabularnewline
\hline
{\small $\tilde{\varphi}\equiv\left(\begin{array}{c}
\tilde{\phi}\\
\tilde{\rho}\end{array}\right)$}&
$\bar\square$&
$\left(\begin{array}{c}
\square\\ {\bf 1} \end{array}\right)$&
$\left(\begin{array}{c}
{\bf 1}\\
\square\end{array}\right)$& $-\frac{1}{2}$ &
$-R$\tabularnewline
\hline
\end{tabular}
\end{center}
\caption{\it The 2-5 Model. We show the ISS matter field decomposition under the gauge $SU(2)$, the flavour $SU(2)_f \times SU(5)_f$ symmetry,
and their charges under the $U(1)_B$ and $R$-symmetry. Both of the $U(1)$ factors above are defined as tree-level symmetries of the
magnetic ISS formulation in Eq.~\eqref{ISSsuperp}. The (small) non-perturbative anomalous effects described by Eq.~\eqref{Wdyn} are not included.
In the absence of baryon-deformations, the $R$-charges of magnetic quarks, $\pm R$, are arbitrary and
can always be re-defined by considering instead a linear combination of $U(1)_B$ and $U(1)_R$ factors.
\label{fieldstableM}}
\end{table}
Note that we use an $f$-suffix
to stand for {}``flavour'' but one should remember that in direct
mediation $SU(5)_{f}$ contains the gauge group of the Standard Model.

In the case of the 2-5 model, by a gauge and
flavour rotation, the matrix $\mu_{ij}^{2}$ can be brought to a diagonal 2-5 form:
\begin{equation}
{\rm 2-5\,\, Model:} \qquad
\mu_{ij}^{2}=\left(\begin{array}{cc}
\mu_{Y}^{2}\mathbf{I}_{2} & 0\\
0 & \mu_{X}^{2}\mathbf{I}_{5}\end{array}\right) \, , \quad \mu_{Y}^{2} > \mu_{X}^{2}
\, .
\label{25mu2}
\end{equation}

Now consider adding the following deformation\footnote{A similar deformation involving a meson operator and two singlet fields
was previously considered in Ref.~\cite{Cho:2007yn}. Their model, however, contained a runaway direction to a supersymmetric vacuum.
For generic values of parameters, this {makes} the non-supersymmetric $R$-breaking vacuum of \cite{Cho:2007yn} short-lived and unstable to
decay in the runaway direction. We will see below that our version of the meson-deformed model
defined by Eqs. \eqref{meson-def}, \eqref{ISSsuperp} with a 2-2-3 or 1-1-5 flavour patterns does not have a supersymmetric
runaway, and the resulting susy-breaking vacuum is stabilised. }
involving the meson
plus some additional singlet fields $A,B,C$:
\begin{equation}
W_{\rm meson-def}=h(m_{1}A^{2}+m_{2}BC+\lambda AB\,\mbox{tr}(\Phi))\,.
\label{meson-def}
\end{equation}
Here we chose to scale all the superpotential parameters with $h$. The meson deformation of the ISS model
is characterised by the dimensionless coupling constant $\lambda$. In the electric-dual ISS formulation this
deformation is $\sim \frac{1}{M_{Pl}} AB \mbox{tr} (Q\tilde{Q})$ and thus
\be
\lambda \sim \frac{\Lambda_{\sst ISS}}{M_{Pl}} \ll 1 \, .
\ee
The new singlet fields are constrained to have $R$-charges given in Table~\ref{ABC-Rs}; these are
different from $0$ or $2$, so spontaneous $R$-symmetry breaking
is a possibility \cite{Shih:2007av,Sun:2008va}.
\begin{table}[h]
\begin{center}
\begin{tabular}{|c|c|}
\hline
 & {\small $U(1)_{R}$}\tabularnewline
\hline
\hline
$A$ & $1$\tabularnewline
\hline
$B$ & $ $$-1$\tabularnewline
\hline
$C$ & $3$\tabularnewline
\hline
\end{tabular}
\end{center}
\caption{\it $R$-charges of $A,B,C$ singlet fields of the meson deformation in Eq.~\eqref{meson-def}.
\label{ABC-Rs}}
\end{table}

The combined effect of $W_{\sst ISS}+W_{\rm meson-def}$, gives a \emph{generic} $R$-symmetry preserving superpotential which defines the
low-energy magnetic formulation of our meson-deformed ISS theory. This is a self-consistent approach since,
as pointed out in Ref.~\cite{Abel:2007nr},
$R$-symmetry breaking
in the electric theory is controlled by a small parameter.\footnote{In principle,
it is known that the apparent $R$-symmetry of the magnetic formulation of the ISS SQCD is an approximate symmetry
of the underlying electric theory: it is broken by the anomaly as per Eq.~\eqref{Wdyn}.
(At the same time, the anomaly-free combination of $U(1)_R$
and the axial symmetry $U(1)_A$ is broken explicitly by the mass terms of electric quarks $m_Q$.)
However, the $R$-symmetry is broken in the electric theory in a controlled way \cite{Abel:2007nr}
by small parameter, $m_Q /\Lambda_{\sst ISS} = \mu^2 /\Lambda_{\sst ISS}^2 \ll 1$. As such the $R$-symmetry is preserved to that order in the
superpotential.}
Terms quadratic in the
meson $\Phi$ that could arise from lower dimensional irrelevant operators
in the electric theory are forbidden by $R$-symmetry.
Thus, our deformation is described by a \emph{generic}
superpotential and $W_{\sst ISS}+W_{\rm meson-def}$ gives its leading-order terms.

Being an exact symmetry of the tree-level magnetic superpotential,
the $R$-symmetry of this model is
actually spontaneously-broken, as we have already alluded to above.
We shall
consider this R-symmetry breaking before we discuss the SUSY breaking and its mediation.

First note that for any non-zero $\langle AB\rangle$ we can define
an effective $\mu^{2}$ term
\begin{equation}
\mu_{\rm eff}^{2}=\mu^{2}-\lambda\langle A \rangle \langle B\rangle.
\end{equation}
Thus the magnetic quarks acquire VEVs precisely as they do in the undeformed
ISS but with $\mu^{2}$ replaced by $\mu_{\rm eff}^{2}$;
\begin{eqnarray}
\langle\rho\rangle & = & \langle\tilde{\rho}\rangle\, =\, 0 \\
\langle\phi\tilde{\phi}\rangle & = & \mu_{\rm Y\,eff}^{2}.
\label{rank-c}
\end{eqnarray}
The VEVs of $\mbox{tr}(\Phi)$ and $C$ will simply set $\langle F_{A}\rangle =\langle F_{B}\rangle =0$;
that is
\begin{eqnarray}
\langle\mbox{tr}(\Phi)\rangle & = & -\frac{2m_{1}\langle A\rangle }{\lambda \langle B\rangle}\label{Phivev} \\
\langle C\rangle & = & -\frac{\lambda \langle A\rangle \, \langle\mbox{tr}(\Phi)\rangle}{m_{2}}
=\frac{2m_{1}\langle A\rangle^{2}}{m_{2}\langle B\rangle}.
\label{Cvev}
\end{eqnarray}
At this point the full potential is
\be
V  =  \sum_{i=3}^{7}h^{2}|(\mu_{\rm eff}^{2})_{ii}|^{2}+|F_{C}|^{2}
 =  5h^{2}|\mu_{X}^{2}-\lambda\langle AB\rangle|^{2}+ h^{2}m_{2}^{2}|B|^{2}\, ,
 \label{V52}
 \ee
so there is a runaway to unbroken SUSY in the direction $B\rightarrow0$
and $A=\mu_{X}^{2}/\lambda B\rightarrow\infty$ along which the $R$-symmetry
is broken.

Now, in order to end up with broken SUSY we would like to stabilize
this type of runaway with Coleman-Weinberg terms in the one-loop potential.
(Note that alternatively one could stabilize the model at tree-level
using a more complicated potential and $R$-symmetry as discussed
in Ref.~\cite{Carpenter:2008wi}.) We therefore need a runaway to
\emph{broken }SUSY since the Coleman-Weinberg contributions vanish
where SUSY is unbroken. The classical runaway vacuum becomes non-supersymmetric
if the components of the $\mu_{X\, ij}^{2}$ matrix on the right hand side of Eq.~\eqref{V52}
are no longer degenerate. This is easily achieved by breaking
the
flavour group into three rather than two factors.

For example, one can consider a 2-2-3 model. Here the original $SU(7)_f$ of the ISS $SU(2)_{\rm mg}$ gauge theory is broken to
$SU(2)_f\times SU(2)_f \times SU(3)_f$. This realisation can be thought of as the 2-5 model above where the $SU(5)_f$ flavour
subgroup was further broken to $SU(2)_f \times SU(3)_f \times U(1)_{\rm traceless}$
by splitting the eigenvalues of the $\mu_{ij}^{2}$ matrix.
This does not cause problems for either explicit or direct mediation.
Indeed in the case of direct gauge mediation the
$SU(2)_{L}$ and $SU(3)_{c}$ components of $\mu_{ij}^{2}$ (or equivalently
$m_{Q}$ in the electric theory) renormalize differently below the
GUT scale and so they are not expected to be the same
\footnote{Note that renormalization of $\mu^{2}$ above the scale $\Lambda_{\sst ISS}$
would be understood as renormalization of $m_{Q}$ in the electric
theory.}.

Alternatively, one can consider an even simpler example of a 1-1-5 model with $N_f=7$ and $N_c=6$ so that the magnetic
`number of colours', $N=1$, and the magnetic group is trivial. By splitting the eigenvalues of the $\mu_{ij}^{2}$ matrix
we choose the flavour breaking to have the 1-1-5 pattern,
$SU(7)_f \rightarrow U(1)_f\times U(1)_f \times SU(5)_f$. For the case of direct mediation the SM gauge group is $SU(5)_f$.

To give a unified treatment of the 1-1-5 and the 2-2-3 models one can consider a general
$N$-$P$-$X$ model with $N+N_P+N_X =N_f$ and the $\mu_{ij}^{2}$ matrix given by:
\begin{equation}
\mu_{ij}^{2}=\left(\begin{array}{ccc}
\mu_{Y}^{2}\mathbf{I}_{N} & 0 & 0\\
0 & \mu_{P}^{2}\mathbf{I}_{N_P} & 0\\
0 & 0 & \mu_{X}^{2}\mathbf{I}_{N_X}\end{array}\right) \, , \quad \mu_{Y}^{2} > \mu_{P}^{2}\, , \, \mu_{X}^{2} \, ,
\quad \mu_{P}^{2} \neq \mu_{X}^{2} \, ,
\label{muYXP}
\end{equation}
which corresponds to $SU(N_f) \rightarrow SU(N)_f\times SU(N_P)_f \times SU(N_X)_f$ as well as traceless $U(1)$ combinations
which commute with the right hand side of Eq.~\eqref{muYXP}.
For simplicity, the rank
of top left $Y$-corner is identified with $N$, the number of magnetic colours, thus
the original ISS rank condition which is responsible for the SUSY-breaking vacuum
is arranged so that $F_{\Phi} =0$ when $\Phi=Y$, see Eq.~\eqref{rank-c},
and $F_{\Phi}  \neq 0$ when $\Phi$ is either $P$ or $X$.
The corresponding decomposition of ISS magnetic matter fields and their charges for this models are given in Table~\ref{fieldstableM2}.
\begin{table}[t]
\begin{center}
{\footnotesize }\begin{tabular}{|c|c|c|c|c|c|}
\hline
$N$-$P$-$X$ Model
 & {\small $SU(N_P)_{f}$} & {\small $SU(N_X)_{f}$} & {\small $SU(N)_{\rm mg}$} & {\small $U(1)_{B}$} & {\small $U(1)_{R}$}\tabularnewline
\hline
\hline
{\small $\Phi_{ij}\equiv\left(\begin{array}{ccc}
Y & N & Z\\
\tilde{N} & P & M\\
\tilde{Z} & \tilde{M} & X\end{array}\right)$} & {\small $\left(\begin{array}{ccc}
1 & \square & 1\\
\bar{\square} & Adj+1 & \square\\
1 & \bar{\square} & 1\end{array}\right)$} & {\small $\left(\begin{array}{ccc}
1 & 1 & \square\\
1 & 1 & \square\\
\bar{\square} & \bar{\square} & Adj+1\end{array}\right)$} & {\small $1$} & {\small $0$} & {\small $2$}\tabularnewline
\hline
{\small $\varphi\equiv\left(\begin{array}{c}
\phi\\
\sigma\\
\rho\end{array}\right)$} & {\small $\left(\begin{array}{c}
1\\
\bar{\square}\\
1\end{array}\right)$} & {\small $\left(\begin{array}{c}
1\\
1\\
\bar{\square}\end{array}\right)$} & {\small $\square$} & {\small $\frac{1}{N}$} & {\small $R$}\tabularnewline
\hline
{\small $\tilde{\varphi}\equiv\left(\begin{array}{c}
\tilde{\phi}\\
\tilde{\sigma}\\
\tilde{\rho}\end{array}\right)$} & {\small $\left(\begin{array}{c}
1\\
\square\\
1\end{array}\right)$} & {\small $\left(\begin{array}{c}
1\\
1\\
\square\end{array}\right)$} & {\small $\bar{\square}$} & {\small $-\frac{1}{N}$} & {\small $-R$}\tabularnewline
\hline
\end{tabular}
\end{center}
\caption{\it The $N$-$P$-$X$ Model. We indicate ISS matter field decomposition under the flavour subgroup
$SU(N_P)_f \times SU(N_X)_f$.
In direct mediation we would gauge $SU(N_P)_f \times SU(N_X)_f\times U(1)_{\rm traceless}$
or its subgroup, and identify it with the SM gauge group.
We also show the gauge $SU(N)$ and the charges under the $U(1)_B$ and $R$-symmetry as in Table~\ref{fieldstableM}.
\label{fieldstableM2}}
\end{table}

The minimization with respect to $C$ and $\mbox{tr}(\Phi)$ are as in Eqs.~\eqref{Phivev}-\eqref{Cvev}
before, but minimization with respect to $A$, results in
\begin{equation}
\langle A\rangle=\,\frac{N_P \mu_P^2+N_X \mu_X^{2}}{N_P+N_X}\, \frac{1}{ \lambda \langle B \rangle }\, ,
\end{equation}
 and consequently the potential
\begin{eqnarray}
V  &=&  \sum_{i=N+1}^{N_f}h^{2}|(\mu_{\rm eff}^{2})_{ii}|^{2}+|F_{C}|^{2} \nonumber \\
& = & h^2 N_P \left(\mu_P^2-\frac{N_P \mu_P^2+N_X \mu_X^{2}}{N_P+N_X}   \right)^2
+ h^2 N_X \left(\mu_X^2-\frac{N_P \mu_P^2+N_X \mu_X^{2}}{N_P+N_X} \right)^2
+ h^2 m_{2}^{2}|B|^{2}
\nonumber \\
 & = & h^{2}\frac{N_P N_X}{N_P+N_X}(\mu_{X}^{2}-\mu_{P}^{2})^{2}+ h^2 m_{2}^{2}|B|^{2}.
 \label{Vclasmes}
 \end{eqnarray}
Again there is a runaway but now to broken supersymmetry as desired.

Note that in the case of explicit mediation the flavour symmetries in the ISS sector are divorced from the
gauge symmetries of the Standard Model. In that case one can have a
breaking of flavour symmetry
that is more general than Eq.~\eqref{muYXP}, in terms of $\mu_{ii}^2 $.
Defining the average $\mu_{ii}$ of the unbroken $SU(N_f-N)$ factor as
\be
\overline{\mu^2}= \frac{1}{N_f-N} \sum_{i=N+1}^{N_f} \mu_{ii}^2\, ,
\ee
we have
\be
\langle A \rangle = \frac{\overline{\mu^2}}{\lambda \langle B \rangle }
\ee
and then the generalisation of Eq.~\ref{Vclasmes} reads
\be
V=h^2 \sum_{i=N+1}^{N_f} (\mu_{ii}^2-\overline{\mu^2} )^2 + h^2 m_{2}^{2}|B|^{2} \, .
\ee
It is worth re-emphasizing that even in the limit $A,C\rightarrow\infty$ and $B\rightarrow 0$ the scalar potential $V$ is non-zero, so we have a
runaway to \emph{broken} SUSY.
Proceeding to one-loop, the Coleman-Weinberg contribution to the
potential is therefore expected to lift and stabilize this direction at the same time as lifting the
pseudo-Goldstone modes.

The Coleman-Weinberg effective potential \cite{Coleman:1973jx} sums up one-loop quantum
corrections into the following form:
\be
V_{\mathrm{eff}}^{(1)}\!=\!\frac{1}{64\pi^2}\,\mathrm{STr}\,{\cal M}^4\log\frac{{\cal M}^2}{
\Lambda^2}\,
\!\equiv\frac{1}{64\pi^2}\left( \Tr\, m_{\rm sc}^4\log\frac{m_{\rm sc}^2}{\Lambda^2}-2\,\Tr\,
  m_{\rm f}^4\log\frac{m_{\rm f}^2}{\Lambda^2} +3\, \Tr\,
m_{\rm v}^4\log\frac{m_{\rm v}^2}{\Lambda^2}\right)\label{CW}
\ee
where $\Lambda$ is the UV cutoff\footnote{Which is traded
for a renormalization scale at which the couplings are defined.},
and the scalar, fermion and vector mass matrices are given
by~\cite{Barbieri:1982nz}:

\be
m_{\rm sc}^2=
\begin{pmatrix}
W^{ab}W_{bc}+D^{\alpha a}D^\alpha_{\phantom{\alpha}c}+D^{\alpha
a}_{\phantom{\alpha}c}D^\alpha\quad &
W^{abc}W_b+D^{\alpha a}D^{\alpha c}\\
W_{abc}W^b+D^{\alpha}_{\phantom{\alpha}a}D^{\alpha}_{\phantom{\alpha}c} &
W_{ab}W^{bc}+D^{\alpha}_{\phantom{\alpha}a}D^{\alpha c}+D^{\alpha
c}_{\phantom{\alpha}a}D^\alpha
\end{pmatrix}
\ee

\be
m_{\rm f}^2=
\begin{pmatrix}
W^{ab}W_{bc}+2D^{\alpha a}D^\alpha_{\phantom{\alpha}c}\quad
&\quad-\sqrt{2}W^{ab}D^\beta_{\phantom{\beta}b}\\
-\sqrt{2}D^{\alpha b}W_{bc} & 2D^{\alpha c}D^\beta_{\phantom{\beta}c}
\end{pmatrix}
\qquad
m_{\rm v}^2=D^\alpha_{\phantom{\alpha}a}D^{\beta a}+D^{\alpha
a}D^\beta_{\phantom{\beta}a}.
\ee
As usual, $W_c\equiv \partial W/\partial \Phi^c = F^\dagger_{\Phi^c}$ denotes a derivative of the
superpotential with respect to the scalar component of the superfield $\Phi^c$ and the raised indices
denote Hermitian conjugation, i.e. $W^{ab} = (W_{ab})^\dagger$.
The $D$-terms are $D^\alpha = g z_a T^{\alpha a}_{\phantom{\alpha}b}z^b$ and they can be formally
switched off by setting the gauge coupling $g=0$, which we shall do for simplicity.
All the above mass matrices will generally depend on
field expectation values.
The effective potential \mbox{$V_{\rm{eff}}=V+V^{(1)}_{\rm{eff}}$} is the sum of the $F$-term (tree-level) potential
and the Coleman-Weinberg contributions.
To find the vacua of the theory we now have to minimize $V_{\rm{eff}}$.

Now
we can check the lifting of
the classical runaway direction by quantum effects in the
Coleman-Weinberg potential.
We have done this numerically using {\em{Mathematica}} and have also checked it with {\em{Vscape}} program of Ref.~\cite{van den Broek:2007kj}.
The non-supersymmetric vacuum is stabilised and in Table~\ref{tabvev115a}
we give values of the VEVs for the 1-1-5 meson-deformed ISS model for a specific choice of external parameters.
It is worth noting at this point that all the
tree-level relations we have just derived get slightly shifted by the one-loop
minization. As we shall see, these one-loop effects often give the leading
contribution to the mediation of SUSY-breaking and
so it is important to keep track of them. This is shown in
Table~\ref{tabvev115a} where in the generic $N$-$P$-$X$ model VEVs develop along the direction
\bea
\vev{\tilde\phi} &=& \xi\,\mathbf{I}_{N}\quad\quad\quad\quad\,\,\vev{\phi}=\kappa\,\mathbf{I}_{N}\nonumber \\
\label{yvevs0}
\vev{Y} &=& \eta\,\mathbf{I}_{N}\quad\quad\quad\quad\vev{P}=p\,\mathbf{I}_{N_P}
\quad\quad\quad\quad\vev{X}=\chi\,\mathbf{I}_{N_X},
\eea
 accompanied by the $A$, $B$, $C$ VEVs as before.
These are the most general VEVs consistent with the tree-level minimization.
\begin{table}[t]
\begin{center}
\begin{tabular}{|c|c|c|c|c|c|c|c|}

\hline
Vev &  $\kappa/\mu_{X}=\xi/\mu_{X}$ &  $\eta/\mu_{X}$& $p/\mu_{X}$ & $\chi/\mu_{X}$& $A/\mu_{X}$ & $B/\mu_{X}$ & $C/\mu_{X}$\tabularnewline
\hline
\hline
Tree-level constrained & $4.7610$ & $0$ & $-0.1283$ & $-4.8242$ & $30.7086$ & $7.5983$ & $248.22$ \tabularnewline
Unconstrained & $4.7607$ & $0.0026$ & $-0.1129$ & $-4.8283$  & $30.7973$ & $7.5617$ & $248.96$ \tabularnewline
\hline
\end{tabular}\par
\end{center}
\caption{\it The 1-1-5 Model: Stabilized VEVs for a meson-deformed ISS theory with
$N_f=7$, $N_c=6$, $h=1$, $m_1/\mu_X=m_2/\mu_X=0.03$, $\mu_Y/\mu_X=5$, $\mu_P/\mu_X=3$ and
$\lambda=0.01$. We
show both the {\em constrained} VEVs (i.e. the VEVs obtained
when the tree-level relations are enforced) and the true
{\em unconstrained} VEVs resulting from complete minimization.
}\label{tabvev115a}
\end{table}

\section{Models of Mediation: from the meson-deformed ISS to the Standard Model} \label{sec:meson-med}

In the context of gauge mediation one can consider two distinct scenarios of how supersymmetry and $R$-symmetry breaking is transmitted to the visible Standard Model sector. The first class is ordinary gauge mediation (i.e. mediation with explicit messengers),
and the second class involves the models of direct gauge mediation.
In this section we {discuss how} these two possibilities can be
realized for the SUSY breaking models we have outlined in the previous section

\subsection{Gauge mediation with explicit messengers} \label{sec:meson-ex}

We begin in this subsection with explicit mediation.
In this scenario one imagines that there is a third sector --
the messenger fields -- that
is responsible for generating the SUSY breaking operators required in the
visible sector.  The approach in this paper is to try to have
a preserved $R$-symmetry that is broken spontaneously. What we shall find is that
we fall foul of the tachyonic messenger problem: ultimately we have to
reintroduce explitcit $R$-symmetry breaking messenger masses to avoid this and we
are forced back to the explicit mediation scenario of Ref.~\cite{Murayama:2006yf}.

To show this, let is first introduce an additional set of mediating fields $f$ and $\tilde{f}$ transforming in the fundamental
(and antifundamental respectively) {of the Standard Model} gauge groups. For concreteness we can take $f$ and $\tilde{f}$
to be (anti)-fundamentals of the underlying GUT gauge group, e.g. $SU(5)_{\rm GUT}$.
In explicit medation these messengers couple to the ISS sector via
additional messenger coupling in the superpotential
\be
\label{Wmess}
W_{\rm mess}=\, \Tr(\tau \Phi)\, f  \cdot \tilde{f} \, ,
\ee
where $\tau_{ij}$ is an arbitrary coupling which from the electric theory perspective
should scale as  $\Lambda_{\sst ISS}/M_{Pl}$ as in Ref.~\cite{Murayama:2006yf}. We remind the
reader that there are no constraints on this coupling coming from the
Standard Model, and that the ISS parameters, such as $N$, $N_f$ are essentially unconstrained.

In order to see how the SUSY breaking enters the visible sector we
need to exhibit the mass matrices for messenger fields explicitly. At tree-level the
SUSY breaking enters into the scalar mass-squared matrices through the
non-zero $F_\Phi$-terms to which the messenger fields, $f$ and $\tilde{f}$ couple.
In general the matrices are given by (ignoring the $D$-terms)
\begin{eqnarray}
m_{\rm sc}^{2} & = & \left(\begin{array}{cc}
W^{ab}W_{bc} & W^{abc}W_{b}\\
W_{abc}W^{b} & W_{ab}W^{bc}\end{array}\right)\, ,\nonumber \\
m_{\rm f} & = & W_{ab}\, ,
\end{eqnarray}
with the $W_{ac}$ being the SUSY preserving mass of the fermions, and
the off-diagonal terms $W^{abc}W_{b}$ containing the SUSY breaking.
In this case $W_{f\tilde{f}}=\Tr(\langle \tau \Phi \rangle )$ is the Dirac mass of
the fermionic superpartners, $\psi_f$ and $\psi_{\tilde{f}}$, and
the SUSY breaking contribution appears first in the tree-level
mass-squared of the scalars, $S=(f,\tilde{f}^*)$.
We have:
\be
m_{\rm sc}^{2}  =  \left(\begin{array}{cc}
| \Tr(\tau \Phi) |^2  & \Tr(\tau^\dagger F_{\Phi}^\dagger) \\
\Tr(\tau F_{\Phi}) & | \Tr(\tau \Phi) |^2 \end{array}\right)\, .
\ee
Now, in order to avoid tachyonic messengers we must here impose
the usual explicit mediation constraint that
\be
| \Tr(\tau \langle \Phi \rangle ) |^2 > |\Tr(\tau \langle F_{\Phi}\rangle)|
\ee
which is effectively a {\em lower} bound on the amount of spontaneous $R$-symmetry
breaking (since $\langle \Phi \rangle$ is charged under $R$-symmetry). In particular this generally prevents us arranging a
split scenario with gauginos much lighter than squarks and sleptons, since this would be a
signature of approximate $R$-symmetry.
(The situation is drastically different in models of direct mediation as we shall see in the following sections.)

As we have said dimensional arguments give \[\tau\sim\lambda\sim \Lambda_{\sst ISS}/M_{Pl}\ll \, 1\]
so the tachyonic inequality is delicate.
If one assumes that $\Phi\sim \mu$
then it seems that the inequality is actually always violated
when $\tau \ll 1$. But note that the same inequality can be equivalently
written in terms of singlet VEVs,
\be
\tau \Phi \sim \tau m_1 \frac{A}{\lambda B} \sim \frac{\tau}{\lambda^2}
\frac{m_1 \mu^2}{B^2 }\, ,
\ee
which shows that the situation is quite complicated and
can only be analyzed numerically. For the values in Table~\ref{tabvev115a}
taking $\tau \sim \lambda$ violates the inequality which suggests that
it may be problematic in general to avoid tachyonic messengers.

An explicit $R$-breaking mass term is a way to overcome
this tachyon so that,
as in Ref.~\cite{Murayama:2006yf}, Eq.~\eqref{Wmess} becomes
\be
\label{Wmess2}
W_{\rm mess}=\, \Tr(\tau \Phi)\, f  \cdot \tilde{f} + M_f\, f  \cdot \tilde{f} \,
\ee
Hence explicit gauge mediation and spontaneous $R$-symmetry breaking are
inconsistent when the DSB is based on the ISS model. Note that
we could have also added a term $\frac{A^2 f  \cdot \tilde{f}}{M_{Pl}}$; however since we have
$\langle A \rangle \sim \mu_P \ll \Lambda$ the effective mass that this induces for the messengers
is even smaller than $\langle \Tr(\tau \Phi)\rangle $.

{}From here on the calculation of the SUSY spectrum is rather standard
with values for gaugino masses being generated being of the same order as those for scalar masses; and so one expects a similar
phenomenology to normal explicit gauge mediation \cite{Giudice:1998bp},
with the diagram that induces the gaugino mass in the present explicit
mediation case as shown in Fig.~\ref{gaugino-expl}.
\begin{figure}[t]
\begin{center}
\begin{picture}(200,120)
\SetOffset(0,20)
\Gluon(0,0)(50,0){4.5}{5}
\ArrowLine(50,0)(0,0)
\Line(50,0)(120,0)
\Gluon(120,0)(170,0){4.5}{5}
\Vertex(50,0){3}
\Vertex(120,0){3}
\DashCArc(85,0)(35,0,180){3}
\ArrowLine(120,0)(170,0)
\Vertex(85,35){3}
\Text(85,50)[c]{\scalebox{1.1}[1.1]{$\Tr (\tau\langle F_{\Phi}\rangle)$}}
\Vertex(85,0){1}
\Line(80,5)(90,-5)
\Line(90,5)(80,-5)
\Text(85,-15)[c]{\scalebox{1.1}[1.1]{$\Tr (\tau \langle \Phi \rangle)$}}
\end{picture}
\end{center}
\vspace*{-0.5cm} \caption{\it One-loop contribution to the
gaugino masses from messengers $f$, $\tilde{f}$. The dashed (solid) line is a bosonic (fermionic) messenger.
The blob on the scalar line indicates an insertion of the $F$-term VEV into the propagator
of the scalar messengers and the cross denotes an
insertion of the $R$-symmetry breaking VEV into the propagator of the fermionic messengers.}
\label{gaugino-expl}
\end{figure}
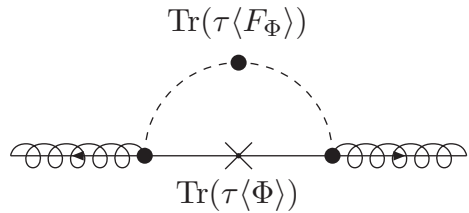

However there is one feature of the present set-up that is rather interesting.
The SUSY breaking effects in the visible sector, i.e. the gaugino
and squark
masses, are all proportional to the combination
$W^{abc}W_{b}=Tr(\tau^\dagger F^\dagger_{\Phi})$. But as we have seen in the previous section,
the $F$-terms at the
minimum (with VEV-less messengers, so that the SM gauge groups are not Higgsed) are given at tree-level by
\be
F^\dagger_{\Phi_{ij}}=h \delta_{ij} (\mu^2_{ii} - \overline{\mu^2}) \, ,
\ee
which clearly obeys
\be
\Tr(F^\dagger_{\Phi})=0 \, .
\ee
This can be seen to result from the minimization of the tree-level
potential with respect to $A$ for a given $B$ VEV:
\be
\frac{\partial V}{\partial A}=\lambda B \Tr(F_\Phi^\dagger)=0\, .
\ee
Thus (at tree-level) the mediation of SUSY-breaking to the visible sector
requires non-degenerate couplings $\tau_{ii}$, and
indeed we can write
\be
\Tr(\tau F_{\Phi}) =
h ( {\overline{\tau \mu^2}} - \bar{\tau} \overline{\mu^2} ) \, .
\ee
That is, only if {\em both} $\tau $ and $\mu$ have non-degeneracy
can there be unsuppressed SUSY breaking mediation, even though
SUSY breaking {\em per se} requires non-degeneracy only in the latter.

However, as we have said, when the full minimization is performed, tree-level
relations such as $\Tr(F^\dagger_{\Phi})=0 $ are no longer expected to hold
(for example, with the unconstrained values in the table we {find
$\Tr(F^\dagger_{\Phi})=-0.034 \mu^{2}_{X} $):} typically one
finds $\Tr(F^\dagger_{\Phi})=\mu^2/(16\pi^2) $,
since the effective $F$-term for mediation is one-loop suppressed.
Thus when the $\tau$ are degenerate one can still
get $m_{\lambda}\sim \frac{\mu^2}{16\pi^2 M_f}\frac{g^2}{(16\pi^2)} \sim 1$~TeV if
$\mu^2/M_f\sim 10^7$~GeV.

\subsection{Direct gauge mediation} \label{sec:meson-dir}

{Now, let us compute gaugino masses} for the direct gauge mediation scenario from the meson-deformed ISS sector.
We first consider the effects {of those direct messengers which obtain
$R$-symmetry breaking masses at tree-level and which couple directly to the
largest $F$-terms.
These transform in the fundamental representation of the SM gauge groups, and
this constitutes a strictly one-loop and formally leading order effect.
Then we will include additional, formally higher-loop, contributions
from the pseudo-Goldstone modes transforming
in both adjoint and (bi-)fundamental representations of the Standard Model gauge groups. It will turn out that the latter contributions can
be of the same order.}

\subsubsection{Strict one-loop contributions to gaugino masses} \label{sec:meson-med-f}

To present a general discussion relevant for any deformation of the ISS model, by mesons, baryons or otherwise,
we shall consider models of the form
\begin{equation}
W=h\Phi_{ij}\varphi_{i}.\tilde{\varphi_{j}}-h\mu_{ij}^{2}\Phi_{ji}+W_{\rm meson-def}(A_{a},\Phi)+
W_{\rm baryon-def}(A_{a},\phi, \tilde{\phi})
\end{equation}
where $A_{a}$ denote generic singlets. The superpotential depends on $\Phi$
linearly, this is dictated by the $R$-symmetry
of the model and is a central feature of direct mediation in the ISS
context.

To keep the presentation simple
in what follows we shall concentrate here on the
1-1-5 model, so that the
parent gauge symmetry of the SM (in this case $SU(5)_f$) is non-split. This discussion can also be straightforwardly generalised to the
2-2-3 and other $N$-$P$-$X$ models by an appropriate reassembling of building blocks below.

The all important messenger/SUSY-breaking
coupling in the superpotential is in this class of models is
\begin{equation}
\frac{1}{h}\, W\, \supset\,  \Phi_{ij}\varphi_{i}.\tilde{\varphi_{j}} \supset
\rho X\tilde{\rho}+\phi Z\tilde{\rho}+\rho\tilde{Z}\tilde{\phi}+\phi Y\tilde{\phi}\,.\label{eq:mess susy breaking}
\end{equation}
The field $\Phi$ is the pseudo-Goldstone mode, although note that
$F_{\phi}$ and $F_{\tilde{\phi}}$ are non-zero as well as $F_{\Phi}$
-- this will be important in what follows.

Gaugino masses are generated at one-loop order as indicated in Fig. \ref{gauginofig}. The fields propagating in the loop are
fermion and scalar components of the direct mediation `messengers'.
Since
gaugino
masses are forbidden by $R$-symmetry one crucial ingredient in their generation is the presence of
non-vanishing $R$-symmetry breaking VEVs.
We are at this point interested in the contribution to the
gaugino mass coming from those messenger fields transforming in the fundamental
of $SU(5)$, which formally give the leading-order contribution. (We shall consider
the contribution from the $X$ fields separately in Section \ref{sec:meson-dir-a}.)
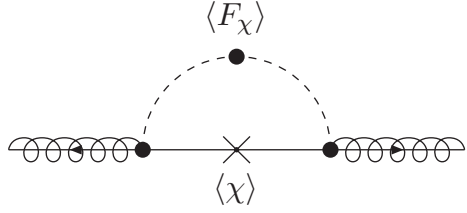
\begin{figure}[t]
\begin{center}
\begin{picture}(200,120)
\SetOffset(0,20)
\Gluon(0,0)(50,0){4.5}{5}
\ArrowLine(50,0)(0,0)
\Line(50,0)(120,0)
\Gluon(120,0)(170,0){4.5}{5}
\Vertex(50,0){3}
\Vertex(120,0){3}
\DashCArc(85,0)(35,0,180){3}
\ArrowLine(120,0)(170,0)
\Vertex(85,35){3}
\Text(85,50)[c]{\scalebox{1.1}[1.1]{$\langle F_{\chi}\rangle$}}
\Vertex(85,0){1}
\Line(80,5)(90,-5)
\Line(90,5)(80,-5)
\Text(85,-15)[c]{\scalebox{1.1}[1.1]{$\langle \chi\rangle$}}
\end{picture}
\end{center}
\vspace*{-0.5cm} \caption{\it One-loop contribution to the
gaugino masses. The dashed (solid) line is a bosonic (fermionic) messenger.
The blob on the scalar line indicates an insertion of $\langle F_{\chi}\rangle$ into the propagator
of the scalar messengers and the cross denotes an
insertion of the $R$-symmetry breaking VEV into the propagator of the fermionic messengers.}
\label{gauginofig}
\end{figure}

First we exhibit the mass matrices of messenger fields. As before, they
are given by (ignoring the $D$-terms)
\be
m_{\rm sc}^{2}  =  \left(\begin{array}{cc}
W^{ab}W_{bc} & W^{abc}W_{b}\\
W_{abc}W^{b} & W_{ab}W^{bc}\end{array}\right)\, , \qquad
m_{\rm f}  = W_{ac}\, .
\ee
The fundamental messengers are $\rho$, $\tilde{\rho}$ and $Z$,
$\tilde{Z}$: we may define a messenger fermion multiplet,
\begin{eqnarray}
\psi & = & (\rho_{i}\,,Z_{i})_{\rm ferm}\, ,\nonumber \\
\tilde{\psi} & = & (\tilde{\rho}_{i}\,,\tilde{Z}_{i})_{\rm ferm}\, ,
\end{eqnarray}
where $i=1..5$. Then $\mathcal{L}\supset\psi\, m_{\rm f}\tilde{\psi}^{T}$
where the fermion messenger mass matrix is
\begin{equation}
m_{\rm f}=\mathbf{I}_{5} \otimes\left(\begin{array}{cc}
\chi & \xi\\
\kappa & 0\end{array}\right),
\end{equation}
written in terms of the VEVs $\chi$, $\kappa$ and $\xi$ (c.f. \eqref{yvevs0}):
\begin{equation}
\langle X\rangle=\chi I_{5}\qquad , \qquad
\langle\phi\rangle=\kappa \qquad , \qquad
\langle\tilde{\phi}\rangle=\xi \, .
\end{equation}
For the scalar mass-squared matrix, we can define equivalent multiplets
\begin{equation}
\label{Sdef}
S=(\rho_{i},Z_{i},\tilde{\rho}_{i}^{*},\tilde{Z}_{i}^{*})_{\,\rm sc}\, .
\end{equation}
To proceed one can diagonalise the mass matrices and compute the
full one-loop contribution to the gaugino mass. That is we define
the diagonalisations:
\begin{eqnarray}
\hat{m}_{\rm sc}^{2} & = & Q^{\dagger}m_{\rm sc}^{2}Q
\label{QdQdef} \\
\hat{m}_{\rm f} & = & U^{\dagger}m_{\rm f}V
\label{UdVdef}
\end{eqnarray}
with eigenvectors
\begin{eqnarray}
\hat{S} & = & S.Q\nonumber \\
\hat{\psi}_{+} & = & \psi.U\nonumber \\
\hat{\psi}_{-} & = & \tilde{\psi}.V^{*}
\label{diagbasis}
\end{eqnarray}
{Here,} the $m_{\rm f}$ diagonalisation is in general a biunitary
transformation.

In order to calculate the gaugino mass, we need the gauge interaction
terms given by
\begin{eqnarray}
\label{gaugeint}
\mathcal{L} & \supset & i\sqrt{2}g_{A}\lambda_{A}(\psi_{1}T^{A}S_{1}^{*}+\psi_{2}T^{A}S_{2}^{*}
+\tilde{\psi}_{1}T^{*A}S_{3}+\tilde{\psi}_{2}T^{*A}S_{4})+H.C. \\
\label{gaugeint2}
 & = & i\sqrt{2}g_{A}\lambda_{A}(\hat{\psi}_{+i}\hat{S}_{k}^{*}(U_{i1}^{\dagger}Q_{1k}+U_{i2}^{\dagger}Q_{2k})+
 \hat{\psi}_{-i}\hat{S}_{k}(Q_{k3}^{\dagger}V_{1i}+Q_{k4}^{\dagger}V_{2i}))+H.C.
 \end{eqnarray}
Then the diagram in Figure \ref{gauginofig} amounts {to\footnote{{More precisely, there are actually two diagrams of this type which are mirror images of each other.}}}
\begin{eqnarray}
M_{\lambda_{A}}^{(\rho,Z)} &\!\!\!=\!\!\!&4 N g_{A}^2\, \tr(T^{A}T^{B})\sum_{ik}(U_{i1}^{\dagger}Q_{1k}
+U_{i2}^{\dagger}Q_{2k})(Q_{k3}^{\dagger}V_{1i}+Q_{k4}^{\dagger}V_{2i})\,
I(\hat{m}_{{\rm f},i},\hat{m}_{{\rm sc},k})
\label{eq:master}
\end{eqnarray}
where $I(\hat{m}_{{\rm f}},\hat{m}_{{\rm sc}})$ is the appropriate one-loop integral with
a fermion and a scalar. Here the ``$N$'' reinstates the possibility of an
$SU(N)_{mg}$ gauge group. In the diagonal mass-basis
\begin{equation}
I(a,b)\!=\!\int \frac{d^{4}k}{(2\pi)^{4}} \frac{a}{k^{2}-a^{2}}\frac{1}{k^{2}-b^{2}}
=\frac{-a(\eta+1)}{16\pi^2}+\frac{1}{16\pi^2}
\frac{a}{a^2-b^2}\left[a^2\log\left(\frac{a^{2}}{\Lambda^{2}}\right)-b^2\log\left(\frac{a^{2}}{\Lambda^{2}}\right)\right]
\label{intIab}
\end{equation}
and
\begin{equation}
\eta=\frac{2}{4-D}+\log(4\pi)-\gamma_{E}.
\end{equation}
This integral is UV-divergent, but the divergences cancels in the sum over eigenstates as required.

Using \eqref{eq:master} we can now evaluate gaugino masses in Figure \ref{gauginofig}
generated by fundamental messengers $\rho$, $\tilde{\rho}$ and $Z$. Numerical values
for the gaugino mass for a few different values of parameters of the model are given in the Tables
in section \ref{sec:meson-sum}.

It is instructive to complement these numerical calculations by a simple analytic estimate, and in particular
explain the smallness of these gaugino mass contributions.
When the $F$-terms are small compared to $\mu^{2}$ one can expand Eqs.~\eqref{eq:master}-\eqref{intIab}.
We define a matrix of `weighted' $F$-terms as:
\be
{\cal F}^{ab} =W^{abc}W_{c}\, ,
\label{Fmatrix}
\ee
and to the leading order in ${\cal F}$ obtain,
\begin{equation}
\label{gaugino-one}
M_{\lambda_{A}} = \frac{g_{A}^2}{8\pi^2}\,N\, \tr(T^{A}T^{B})\, \Tr({\cal F} \cdot m^{-1}_{\rm f}) + {\cal O}({\cal F}^3) \, .
\end{equation}
This is a well-known leading order in ${\cal F}$ approximation which is basis-independent. In the Appendix
we give the derivation of Eq.~\eqref{gaugino-one} in the general settings relevant to our model(s).

Clearly the matrix ${\cal F}$ is determined entirely by the contribution
in Eq.~(\ref{eq:mess susy breaking}) to be
\begin{equation}
{\cal F}=W^{abc}W_{c}=h \left(\begin{array}{cc}
 F_{\chi} &  F_{\tilde{\phi}}\\
 F_{\phi} & 0\end{array}\right)
\end{equation}
 and since $m_{\rm f}^{-1}=\left(\begin{array}{cc}
0 & \frac{1}{\kappa}\\
\frac{1}{\xi} & -\frac{\chi}{\xi\kappa}\end{array}\right)$ we find
\begin{equation}
M_{\lambda_{A}}^{(\rho,Z)}
=\frac{g_{A}^2}{8\pi^2}\,N\,\tr(T^{A}T^{B})\,\left(\frac{F_{\tilde{\phi}}}{\xi}+\frac{F_{\phi}}{\kappa}\right)
+ {\cal O}({\cal F}^3)
\label{eq:lowest order Ma}
\end{equation}
Now consider the minimization condition for the tree-level potential,
$V=\sum_c |F^{c}|^{2}$ with respect to $Y^{*}$.
 \begin{equation}
\frac{1}{2} \frac{\partial V}{\partial Y^{*}}=0= \sum_c W^{Yc}F_{c}=\kappa F_{\tilde{\phi}}+\xi F_{\phi}+W_{\rm meson-def}^{YA_{a}}F_{A_{a}}
\end{equation}
(For the constrained 1-1-5 VEVs shown in Table~\ref{tabvev115a}
this trivially sets $\eta=0$.)
This equation together with Eq.(\ref{eq:lowest order Ma}) implies
that the tree-level leading order gaugino mass is zero
\begin{equation}
M_{\lambda_{A}}^{(\rho,Z)}
= \, 0
+ \,{\cal O}({\cal F}^3)
\label{lead-van}
\end{equation}
unless the additional
singlet fields appearing in the meson deformation have non-zero $F$-terms
as well. (This would require an additional source of SUSY breaking beyond
the O'Raifeartaigh breaking of the ISS sector, and is therefore unattractive.)
As we have stressed, these relations are perturbed
when the potential is stabilized by one-loop effects
(e.g. $\eta $ is non-zero
in the unconstrained model of Table \ref{tabvev115a}):
then the estimate in Eq.(\ref{eq:lowest order Ma})
is still reasonably good, with the $F$-terms being derived from the
one-loop equations.

This leading order suppression for the gaugino mass explains the relative smallness of our numerical results
in Table~\ref{compare115a} which shows the ``reduced gaugino
masses'' $m_{1/2}$ defined by
\be
\label{eq:redmh}
M_{\lambda_A} = \frac{g_A^2}{16\pi^2} m_{1/2}\, .
\ee
In particular these values are much smaller
than those derived for the scalars in Table~\ref{comparescalar115a}
where we show the ``reduced scalar masses'' $m_0$ defined by
\be
\label{eq:redm0}
m_{\rm sferm}^2 = \sum _A \frac{g_A^4}{(16\pi^2)^2} C_A S_A m_0^2\, ,
\ee
where $C_A$ and $S_A$ are the standard Casimir/Dynkin indices as in
Ref.~\cite{Martin:1996zb}. We note that this suppression
is also related to that in Ref.~\cite{Izawa:1997gs},
which tells us that $F_{\Phi}$ does not contribute to the gaugino
masses at leading order because of the structure of $m_{\rm f}$ (in
particular the zero entry). Here we find that the argument extends
to quite general models of direct mediation.

\subsubsection{Additional contributions to gaugino masses} \label{sec:meson-dir-a}

The effects considered above have so far generated rather small contributions to gaugino masses.
Thus, we have to consider additional contributions, due to the
adjoint $X$ and $P$ as well as the bifundamental $M$ and $\tilde{M}$ messengers.
 These messengers are massless at tree-level and acquire masses only at loop-level. Thus their contributions to gaugino masses  are formally a higher-loop effect.
After a careful consideration we find that these indeed give a contribution to the gaugino masses which comparable to the strict one-loop effect described above. Scalar masses being unsuppressed at leading order are not significantly effected.

For 1-1-5 type models where the SM gauge group is $SU(5)_f$, the new contributions arise from the
$X_{ij}$ fields with $i,j=1\ldots5$. They contribute through
the diagram shown in Figure \ref{Xmassfig}.
\begin{figure}[t]
\begin{center}
\begin{picture}(300,220)
\scalebox{1.8}[1.8]{
\SetOffset(0,20)
\Gluon(0,0)(50,0){4.5}{5}
\Text(25,10)[c]{\scalebox{0.9}[0.9]{$\lambda$}}
\Text(145,10)[c]{\scalebox{0.9}[0.9]{$\lambda$}}
\ArrowLine(50,0)(0,0)
\ArrowLine(50,0)(85,0)
\ArrowLine(120,0)(85,0)
\Gluon(120,0)(170,0){4.5}{5}
\Vertex(50,0){3} \Vertex(120,0){3}
\DashCArc(85,0)(35,0,180){3}
\ArrowLine(120,0)(170,0)
\Vertex(85,35){3}
\DashLine(85,35)(100,55){3}
\ArrowLine(62.25,26.75)(58.25,23.25)
\ArrowLine(107.25,26.75)(111.75,23.25)
\DashLine(85,35)(70,55){3}
\Text(100,62)[c]{\scalebox{0.7}[0.7]{$\langle \chi^{\dagger}\rangle$}}
\Text(70,62)[c]{\scalebox{0.7}[0.7]{$\langle \chi^{\dagger}\rangle$}}
\Vertex(85,0){1}
\Line(80,5)(90,-5)
\Line(90,5)(80,-5)
\Text(85,-13)[c]{\scalebox{0.7}[0.7]{$m^{\star}_{\chi}$}}
\Text(75,-5)[c]{\scalebox{0.7}[0.7]{$\chi^{\dagger}$}}
\Text(95,-5)[c]{\scalebox{0.7}[0.7]{$\chi^{\dagger}$}}
\Text(55,-5)[c]{\scalebox{0.7}[0.7]{$\chi$}}
\Text(47,12)[c]{\scalebox{0.7}[0.7]{$\chi^{\dagger}$}}
\Text(73,39)[c]{\scalebox{0.7}[0.7]{$\chi$}}
\Text(97,39)[c]{\scalebox{0.7}[0.7]{$\chi$}}
\Text(127,12)[c]{\scalebox{0.7}[0.7]{$\chi^{\dagger}$}}
\Text(115,-5)[c]{\scalebox{0.7}[0.7]{$\chi$}}}
\end{picture}
\end{center}
\caption{\it One-loop contribution to the
gaugino masses from $X$-messengers. The dashed (solid) line is a bosonic (fermionic)
component of $X$. The blob on the scalar line indicates an insertion of
$\langle F_{\chi}\rangle$ into the propagator of the scalar
messengers and the cross denotes an insertion of the $R$-symmetry
breaking VEV into the propagator of the fermionic
messengers.}\label{Xmassfig}
\end{figure}
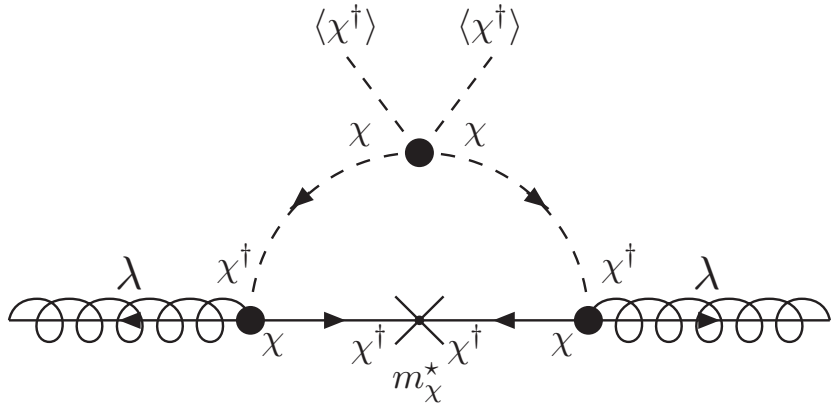
Note that the scalar vertex exists
because the Coleman-Weinberg potential induces an $R$-symmetry violating
mass term. The fermion mass-propagator is also absent at tree-level:
since it is a Majorana term (and the $X$-fermions have $R$-charge
1) it also violates $R$-symmetry and by the non-renormalization theorem
it vanishes in the absence of both $R$-symmetry and supersymmetry
breaking. The naive expectation is therefore that this contribution
will be three-loop suppressed. As we shall see, this is not the case,
and in fact the contribution can be competitive with the previous
contributions. This is because the $X$ modes are pseudo-Goldstone
modes: \emph{all} their masses arise at one-loop, and the lightness
of these modes corresponds to a suppression of the effective messenger
scale of the adjoints whose mass is in fact similar to $M_{SUSY}$.

Let us estimate these effects in more detail. First the mass-insertions:
the scalar mass-squareds come from the Coleman-Weinberg term
\begin{equation}
V_{\mathrm{eff}}^{(1)} \supset\mbox{STr}(\frac{{\cal M}^{4}}{64\pi^{2}}\log {\cal M}^{2}).
\end{equation}
In particular there are terms involving
$\bar{W}_{\rho Z}W^{Z\rho}\bar{W}_{\rho\tilde{\rho}}W^{\tilde{\rho}\rho}=h^{4}\xi^{2}|\delta X_{ij}|^{2}$
where $X=\langle X\rangle+\delta X$. Since typically $\xi\gg\mu\gg\kappa$
one expects $R$-symmetry conserving mass-squared for the adjoints
of order
\begin{equation}
m_{XX^{*}}^{2}\sim\frac{h^{4}\xi^{2}}{64\pi^{2}}
\end{equation}
at the minimum. $R$-symmetry violating masses are induced by terms
such as
$W_{\rho\tilde{\rho}}\bar{W}^{\tilde{\rho}\rho}W_{\rho\tilde{\rho}}\bar{W}^{\tilde{\rho}\rho}\supset h^{4}
\langle X\delta X^{\dagger}\rangle^{2}+h.c=h^{4}\chi^{2}(\delta X_{ij}^{*}\delta X_{ji}^{*})+h.c.$
Hence we expect a neutral mass-squared matrix for $\mathbf{X}=(X^{A},X_{A}^{*})$
(where $A$ is the adjoint index) of the form
\begin{eqnarray}
m_{\mathbf{X}}^{2}& \sim &
\frac{\delta_{AB}}{64\pi^{2}}
\left(\begin{array}{cc}
a & b\\
b^* & a \end{array}\right)\, , \nonumber \\
&& a\sim  \xi^2 \,\,\,\, ; \,\,\, b \sim \chi^2 \, .
\end{eqnarray}
Assuming $b$ is real, the diagonalization of this matrix is
$\hat{m}_{\mathbf{X}}^{2}=({\cal Q}^{X})^{T}m_{\mathbf{X}}^{2}{\cal Q}^{X}=$ \linebreak \mbox{$\frac{h^{4}}{64\pi^{2}}\mbox{diag}(a+b,\,a-b)$}
where
\begin{equation}
\label{QXmatrix}
{\cal{Q}}^{X}=\frac{1}{\sqrt{2}}\left(\begin{array}{cc}
1 & -1\\
1 & 1\end{array}\right).
\end{equation}
We will call the two eigenvalues $\hat{m}_{X_{\pm}}^{2}$.

The $R$-breaking mass term for the adjoint fermion is generated from
diagram shown in Figure \ref{gauginofigX}.
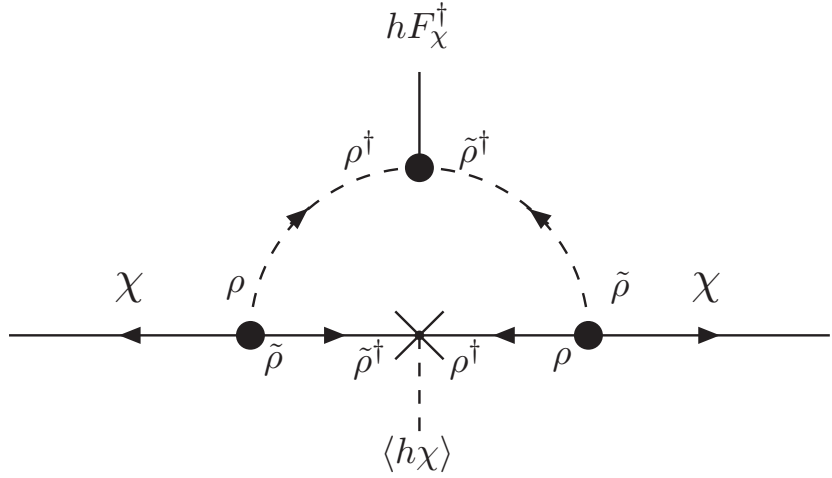
\begin{figure}[t]
\begin{center}
\begin{picture}(300,220)
\scalebox{1.8}[1.8]{
\SetOffset(0,20)
\ArrowLine(50,0)(0,0)
\ArrowLine(50,0)(85,0)
\ArrowLine(120,0)(85,0)
\Text(25,10)[c]{\scalebox{0.9}[0.9]{$\chi$}}
\Text(145,10)[c]{\scalebox{0.9}[0.9]{$\chi$}}
\Vertex(50,0){3} \Vertex(120,0){3}
\DashCArc(85,0)(35,0,180){3}
\ArrowLine(120,0)(170,0)
\Vertex(85,35){3}
\ArrowLine(58.25,23.25)(62.25,26.75)
\ArrowLine(111.75,23.25)(107.25,26.75)
\Text(85,65)[c]{\scalebox{0.7}[0.7]{$hF^{\dagger}_{\chi}$}}
\Vertex(85,0){1}
\Line(80,5)(90,-5)
\Line(90,5)(80,-5)
\Line(85,35)(85,55)
\DashLine(85,0)(85,-20){3}
\Text(85,-25)[c]{\scalebox{0.7}[0.7]{$\langle h\chi\rangle$}}
\Text(75,-5)[c]{\scalebox{0.7}[0.7]{$\tilde{\rho}^{\dagger}$}}
\Text(95,-5)[c]{\scalebox{0.7}[0.7]{$\rho^{\dagger}$}}
\Text(55,-5)[c]{\scalebox{0.7}[0.7]{$\tilde{\rho}$}}
\Text(47,10)[c]{\scalebox{0.7}[0.7]{$\rho$}}
\Text(73,39)[c]{\scalebox{0.7}[0.7]{$\rho^{\dagger}$}}
\Text(97,39)[c]{\scalebox{0.7}[0.7]{$\tilde{\rho}^{\dagger}$}}
\Text(127,10)[c]{\scalebox{0.7}[0.7]{$\tilde{\rho}$}}
\Text(115,-5)[c]{\scalebox{0.7}[0.7]{$\rho$}}}
\end{picture}
\end{center}
\caption{\it One-loop contribution to the
Majorana masses of $X$-fermions. The dashed (solid) line is a bosonic (fermionic)
messenger. The blob on the scalar line indicates an insertion of
$\langle F_{\chi}\rangle$ into the propagator of the scalar
messengers and the cross denotes an insertion of the $R$-symmetry
breaking VEV into the propagator of the fermionic
messengers.}\label{gauginofigX}
\end{figure}
The topology is identical to the one-loop gaugino diagram with internal states
$\psi,\,
\tilde{\psi}$ and $S$
with the mass matrices and diagonalisations as in Eqs.~\eqref{QdQdef} and
\eqref{UdVdef}, although of
course the vertices are different: they come from the $W\supset h\rho X\tilde{\rho}$
coupling and are given by
\begin{equation}
V\supset h\,(X\psi_{1})S_{3}^{*}+h(X\tilde{\psi}_{1})S_{1}\,+h.c.
\end{equation}
In terms of the previous mass eigenstates these become
\[
V\supset hX\,(\hat{\psi}_{+i}\hat{S}_{k}^{*}(U_{i1}^{\dagger}Q_{3k})+\hat{\psi}_{-i}\hat{S}_{k}(Q_{k1}^{\dagger}V_{1i})+H.C.
\]
where the diagonalisation matrices $Q$, $U$ and $V$ are exactly the same as in Eqs.~\eqref{QdQdef}-\eqref{UdVdef}.
Defining $X_{ij}=\sqrt{2}\, X_{A}T_{ij}^{A}$, and with standard
Feynman parametization we find
that the diagram in Figure \ref{gauginofigX} generates
\begin{equation}
\label{mxexp}
M_{\psi_{X}}=4 N h^{2}\tr(T^{A}T^{B})\sum_{ik}(U_{i1}^{\dagger}Q_{3k})(Q_{k1}^{\dagger}V_{1i})\,
I(\hat{m}_{{\rm f},i},\hat{m}_{{\rm sc},k})\, ,
\end{equation}
where $I(\hat{m}_{{\rm f},i},\hat{m}_{{\rm sc},k})$ is the same integral \eqref{intIab} as in \eqref{eq:master}.

Note that although the diagram in Figure \ref{gauginofigX} is similar to the fundamental contribution
to the one-loop gaugino mass, there is less suppression. This is because
the couplings of $\rho,\tilde{\rho}$ and $Z,\,\tilde{Z}$ to $X$
are not degenerate as they are for the gaugino, indeed there is no
equivalent of the $h\rho X\tilde{\rho}$ coupling for the $Z,\tilde{Z}$
fields at all; hence unitarity does not operate in the same way.
Following the same steps as for the gaugino in the Appendix we obtain a non-vanishing leading order result in ${\cal{F}}$,
\begin{equation}
M_{\psi_{X}^{A}}\approx 4\,N\,h^2\tr(T^{A}T^{B}) \sum_{ijk} {\cal A}_{jk}(U_{i1}^{\dagger}V_{1j})(U_{k1}^{\dagger}V_{1i})\,
(\hat{m}_{\rm f})_{i} \, J(\hat{m}_{\rm f}^{2}{}_{i},\,\hat{m}_{\rm f}^{2}{}_{j},\,\hat{m}_{\rm f}^{2}{}_{k})
\label{mxexp2}
\end{equation}
where the matrix ${\cal A}_{ij}$ was defined in Eq.~\eqref{calA}
 and the function $J$ is given by
\begin{equation}
J(a,b,c)=\frac{1}{8\pi^{2}}\frac{a^2b^2\log\left(\frac{a}{b}\right)+a^2c^2\log\left(\frac{c}{a}\right)+b^2c^2\log\left(\frac{b}{c}\right)}
{(a^2-b^2)(a^2-c^2)(b^2-c^2)}.
\end{equation}

A very rough simple estimate is
\begin{equation}
M_{\psi_{X}^{A}}\sim\frac{h^2\chi}{32\pi^{2}}\frac{F_{X}}{\xi^{2}}.
\end{equation}
This should be compared to the equivalent contribution to the gaugino mass in Section~\ref{sec:meson-med-f} which did vanish at this order
(see Eqs.~\eqref{gaugino-one},\eqref{lead-van}).

Having determined the masses of  $X$ messengers we can now make an estimate for their contribution to the gaugino mass.
The general expression is
\bea
M_{\lambda_{A}}^{(X)}&=& g_A^2 \, N_{X}\,\left(I(M_{\psi_{X}},\hat{m}_{X_{+}})-I(M_{\psi_{X}},\hat{m}_{X_{-}})\right) \, ,
\label{eq:masterX}
\eea}
where $N_{X}$ is the rank of the $X$ lower-right corner
in Eq.~\eqref{muYXP}, which in the case of 1-1-5 type models is
$N_{X}=5$.

Equation \eqref{eq:masterX} allows us to evaluate gaugino masses
generated by adjoint $X$-messengers. Numerical values for the full mass expressions (without relying on estimates and expansions in~${\cal{F}}$)
for the model given in Table~\ref{tabvev115a} are presented in Table \ref{compare115a}
in section \ref{sec:meson-sum}. In this table we give contributions from the $\rho$ and $Z$ messengers in the first column and from the $X$
messengers in the second column. The third column gives the similar contribution from $M$ messengers which we will comment on momentarily (see Eq.~\eqref{Mcontrib}).
The last column is the total result. Other tables in the same subsection follow the same structure and give results for other models.

To understand the order of magnitude can also be understood with the help of the following analytical estimates.
As we have seen the masses are of the order
\begin{eqnarray*}
M_{\psi_{X}^{A}} & \sim & \frac{h^{3}\chi}{32\pi^{2}}\frac{\mu^{2}}{\xi^{2}}\\
\hat{m}_{X_{\pm}}^{2} & \sim & \frac{h^{4}}{64\pi^{2}}(\xi^{2}\pm\chi^{2}).
\end{eqnarray*}
Thus for $h\lesssim1$ we expect $M_{\psi_{X}^{A}}^{2}\ll\hat{m}_{X_{\pm}}^{2}$
and we find
\be
M_{\lambda_{A}}^{(X)}\, = \, \frac{g_A^2\, N_{X}}{32\pi^{2}}\,
M_{\psi_{X}^{A}}\log\left(\frac{\hat{m}_{X_{+}}^{2}}{\hat{m}_{X_{-}}^{2}}\right)
\, \sim \, \frac{g_A^2 h^{3}\, N_{X}}{2(16\pi^{2})^{2}}\,\frac{\chi^{3}\mu^{2}}{\xi^{4}},
\label{gauginoX}
 \ee
where the last expression is valid for $\chi\lesssim\xi$. Note that,
although in a ``mass-insertion approximation'' the leading order
diagram is in principle three-loop, there is only a two-loop $1/(16\pi^{2})^{2}$
suppression.

In addition to the contribution from the adjoint $X$ fields we have a contribution from the
$M$ and $\tilde{M}$ fields. As can be seen from Table~\ref{fieldstableM2} these are bifundamentals under the SU($N_X$) and SU($N_P$) groups.
\begin{equation}
M_{\psi_{M}}= N h^{2}\left[\sum_{ik}(U^{P\dagger}_{i1}Q^{X}_{3k})(Q^{X\dagger}_{k1}V^{P}_{1i})\,
I(\hat{m}^{P}_{{\rm f},i},\hat{m}^{X}_{{\rm sc},k})+(X\leftrightarrow P)\right].
\end{equation}
Here, the labels $P$ and $X$ indicate the diagonalization matrizes for the SU($N_{P}$) and SU($N_{X}$) blocks, respectively (see Table~\ref{fieldstableM2}).
In particular, the $X$ is the diagonalization for the $\rho$ and $Z$ messengers whereas $P$ corresponds to the $\sigma$ and $N$.

The corresponding contribution to the gaugino mass is,
\begin{equation}
\label{Mcontrib}
M^{M}_{\lambda^{A}}=4\,\tr(T^{A}T^{B})N_{P}\sum_{k=1}^{2}{\cal Q}^{M}_{1k}({\cal Q}^{M})^{T}_{k2}I(M_{\psi_{M}},\hat{m}_{M,kk}),
\end{equation}
where ${\cal Q}^{M}$ is the $M$-analog of ${\cal Q}^{X}$ matrix given in Eq.~\eqref{QXmatrix}.
As mentioned earlier these contributions are shown in the third column of Table~\ref{compare115a} and similar ones in the subsection~\ref{sec:meson-sum}.

\subsubsection{Scalar masses}\label{meson-scalar}
Having determined
the
gaugino masses in the preceding subsections,
we now outline the procedure for the generation of sfermion masses of the supersymmetric standard model.
As in Ref.~\cite{Abel:2007nr} we follow the calculation of Martin in Ref.~\cite{Martin:1996zb} adapted to our direct mediation models.

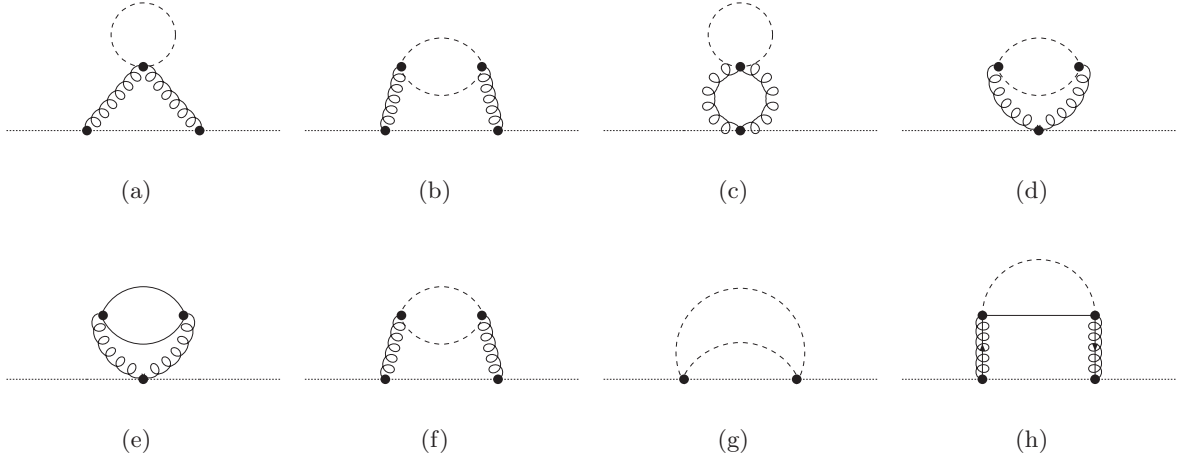
\begin{figure}[t]
\begin{center}
\subfigure[]{
\scalebox{0.6}[0.6]{\begin{picture}(174,120)
\SetOffset(0,20)
\DashLine(0,0)(50,0){1}
\DashLine(50,0)(120,0){1}
\DashLine(120,0)(170,0){1}
\Vertex(50,0){3}
\Vertex(120,0){3}
\Vertex(85,40){3}
\Gluon(50,0)(85,40){4}{6}
\Gluon(85,40)(120,0){4}{6}
\DashCArc(85,60)(20,0,360){3}
\end{picture}}\label{twoloop1}}
\subfigure[]{
\scalebox{0.6}[0.6]{\begin{picture}(174,120)
\SetOffset(0,20)
\DashLine(0,0)(50,0){1}
\DashLine(50,0)(120,0){1}
\DashLine(120,0)(170,0){1}
\Vertex(50,0){3}
\Vertex(120,0){3}
\Gluon(50,0)(60,40){4}{5}
\Gluon(110,40)(120,0){4}{5}
\Vertex(60,40){3}
\Vertex(110,40){3}
\DashCArc(85,30)(28,30,150){3}
\DashCArc(85,50)(28,210,330){3}
\end{picture}}\label{twoloop2}}
\subfigure[]{
\scalebox{0.6}[0.6]{\begin{picture}(174,120)
\SetOffset(0,20)
\DashLine(0,0)(50,0){1}
\DashLine(50,0)(120,0){1}
\DashLine(120,0)(170,0){1}
\Vertex(85,0){3}
\Vertex(85,40){3}
\GlueArc(85,20)(20,90,270){-4}{6.5}
\GlueArc(85,20)(20,270,450){-4}{6.5}
\DashCArc(85,60)(20,0,360){3}
\end{picture}}\label{twoloop3}}
\subfigure[]{
\scalebox{0.6}[0.6]{\begin{picture}(174,120)
\SetOffset(0,20)
\DashLine(0,0)(50,0){1}
\DashLine(50,0)(120,0){1}
\DashLine(120,0)(170,0){1}
\Vertex(85,0){3}
\GlueArc(92,38)(35,175,260){4}{5}
\GlueArc(78,38)(35,280,365){4}{5}
\Vertex(60,40){3}
\Vertex(110,40){3}
\DashCArc(85,30)(28,30,150){3}
\DashCArc(85,50)(28,210,330){3}
\end{picture}}\label{twoloop4}}
\subfigure[]{
\scalebox{0.6}[0.6]{\begin{picture}(174,120)
\SetOffset(0,20)
\DashLine(0,0)(50,0){1}
\DashLine(50,0)(120,0){1}
\DashLine(120,0)(170,0){1}
\Vertex(85,0){3}
\GlueArc(92,38)(35,175,260){4}{5}
\GlueArc(78,38)(35,280,365){4}{5}
\Vertex(60,40){3}
\Vertex(110,40){3}
\CArc(85,30)(28,27,153)
\CArc(85,50)(28,207,333)
\end{picture}}\label{twoloop5}}
\subfigure[]{
\scalebox{0.6}[0.6]{\begin{picture}(174,120)
\SetOffset(0,20)
\DashLine(0,0)(50,0){1}
\DashLine(50,0)(120,0){1}
\DashLine(120,0)(170,0){1}
\Vertex(50,0){3}
\Vertex(120,0){3}
\Gluon(50,0)(60,40){4}{5}
\Gluon(110,40)(120,0){4}{5}
\Vertex(60,40){3}
\Vertex(110,40){3}
\DashCArc(85,30)(28,30,150){3}
\DashCArc(85,50)(28,210,330){3}
\end{picture}}\label{twoloop6}}
\subfigure[]{
\scalebox{0.6}[0.6]{\begin{picture}(174,120)
\SetOffset(0,20)
\DashLine(0,0)(50,0){1}
\DashLine(50,0)(120,0){1}
\DashLine(120,0)(170,0){1}
\Vertex(50,0){3}
\Vertex(120,0){3}
\DashCArc(85,-17)(40,30,150){3}
\DashCArc(85,17)(40,-30,-150){3}
\end{picture}}\label{twoloop7}}
\subfigure[]{
\scalebox{0.6}[0.6]{\begin{picture}(174,120)
\SetOffset(0,20)
\DashLine(0,0)(50,0){1}
\DashLine(50,0)(120,0){1}
\DashLine(120,0)(170,0){1}
\Vertex(50,0){3}
\Vertex(120,0){3}
\Gluon(50,0)(50,40){4}{5}
\ArrowLine(50,0)(50,40)
\Gluon(120,40)(120,0){4}{5}
\ArrowLine(120,40)(120,0)
\Vertex(50,40){3}
\Vertex(120,40){3}
\DashCArc(85,40)(35,0,180){3}
\Line(50,40)(120,40)
\end{picture}}\label{twoloop8}}
\end{center}
\vspace*{-0.5cm} \caption{\small Two-loop diagrams contributing to the sfermion masses.
The long dashed (solid) line is a bosonic (fermionic) messenger.
Standard model sfermions are depicted by short dashed lines.}\label{sfermionfig}
\end{figure}

Sfermion masses are generated by
the
two-loop diagrams shown in Fig. \ref{sfermionfig}.
In \cite{Martin:1996zb} the contribution of these diagrams to the sfermion masses
was
determined to be,
\begin{equation}
m^{2}_{\tilde{f}}=\sum_{mess.} \sum_{a} g^{4}_{a} C_{a} S_{a}(mess.)[{\rm{sum}}\,\,{\rm{of}}\,\,{\rm{graphs}}],
\label{sfermionmass}
\end{equation}
where we sum over all gauge groups under which the sfermion is charged, $g_{a}$ is the corresponding gauge coupling,
$C_{a}=(N^{2}_{a}-1)/(2N_{a})$ is the quadratic Casimir and $S_{a}(mess.)$ is the Dynkin index of the messenger fields
(normalized to $1/2$ for fundamentals).

As in the calculation of the gaugino mass we use the propagators in the diagonal form and insert the
diagonalisation matrices directly at the vertices.
For the diagrams \ref{twoloop1} to \ref{twoloop6} we have closed loops of purely bosonic or purely fermionic
mass eigenstates of our messenger fields.
It is straightforward to check that in this case the unitary matrices from the diagonalisation drop out.
We then simply have to sum over all mass eigenstates the results for these diagrams computed in Ref.~\cite{Martin:1996zb}.

The next diagram \ref{twoloop7} is slightly more involved.
This diagram arises from the D-term interactions. D-terms distinguish between chiral and antichiral fields,
in our case $\rho,Z$ and $\tilde{\rho},\tilde{Z}$,
respectively. We have defined our scalar field $S$ in \eqref{Sdef} such that all component fields have equal charges.
Accordingly, the ordinary gauge vertex is
proportional to a unit matrix in the component space (cf. Eq. \eqref{gaugeint}).
This vertex is then `dressed' with our diagonalisation matrices when we switch to
the $\hat{S}$ basis, \eqref{gaugeint2}.
This is different for diagram \ref{twoloop7}. Here we have an additional minus-sign between chiral and antichiral fields.
In field space this corresponds to
 a vertex that is proportional to a matrix $V_{D}={\rm diag}(1,1,-1,-1)$.
We therefore obtain,
\begin{equation}
{\rm Fig.}~\ref{twoloop7}\, = \, \sum_{i,m} (Q^{T}V_{D}Q)_{i,m}J(\hat{m}_{0,m},\hat{m}_{0,i})(Q^{T}V_{D}Q)_{m,i},
\end{equation}
where $J$ is the appropriate two-loop integral for Fig.~\ref{twoloop7} which can be found in \cite{Martin:1996zb}.

Finally, in \ref{twoloop8} we have a mixed boson/fermion loop.
The subdiagram containing the messengers is similar to the diagram for the gaugino mass. The only difference is the direction of the
arrows on the gaugino lines. Indeed the one-loop sub-diagram corresponds to a contribution
to the kinetic term rather than a mass term for the gauginos.
(The mass term will of course contribute as well but will be suppressed
by quark masses.)
Using Eq. \eqref{gaugeint2} we find,
\begin{eqnarray}
{\rm Fig.}~ \ref{twoloop8} & = & \sum_{ik}(|U_{i1}^{\dagger}Q_{1k}+U_{i2}^{\dagger}Q_{2k}|^{2}
+|Q_{k3}^{\dagger}V_{1i}+Q_{k4}^{\dagger}V_{2i}|^{2})L(\hat{m}_{1/2,i},\hat{m}_{0,k}^{2})\,,
\end{eqnarray}
where $L$ is again the appropriate loop integral from \cite{Martin:1996zb}.

Summing over all diagrams we find the sfermion masses which are typically significantly larger than the gaugino masses calculated earlier.
Indeed, the scalar masses roughly follow the estimate
\begin{equation}
m^{2}_{\tilde{f}}\sim \frac{g^4}{(16\pi^2)^{2}} \mu^{2}.
\end{equation}
This is precisely the leading order effect which in our direct mediation scenario is absent for the gaugino masses.

So far we have taken into account the $\rho,Z$ (or similarly the $\sigma,M$) contributions which as we just explained give
a non-vanishing leading order effect. In distinction to our earlier calculation of the gaugino masses we do not need to include the sub-dominant
contributions from other messengers (which were massless at tree-level)\footnote{Inclusion of such effects would be actually not completely straightforward
because our mass-insertion technique breaks down when used in the two-loop diagrams for the scalars. The reason
for this can be traced to the non-cancelation of the UV cutoff dependent terms. This problem would disappear if one performs a complete higher-loop calculation.
In any case since the leading order result for scalars was non-vanishing we do not expect any significant changes from this.}.

\subsection{Summary of signatures in the directly mediated meson-deformed model} \label{sec:meson-sum}

Here we present and summarize our result for gaugino and sfermion masses for a variety of our meson-deformed models.
These results are most conveniently expressed in terms of the reduced gaugino ($m_{1/2}$)
\be
\label{gaugino1}
M_{\lambda_A} = \frac{g_A^2}{16\pi^2} m_{1/2}\, ,
\ee
and scalar masses ($m_0^2$)
\be
\label{sferm1}
m_{\rm sferm}^2 = \sum _A \frac{g_A^4}{(16\pi^2)^2} C_A S_A m_0^2.
\ee
We similarly define reduced masses for the pseudo-Goldstone components of the direct messengers (appearing in Tables~\ref{comparespectrum115a}, \ref{comparespectrum115b},
\ref{comparespectrum223})
by including a factor of $16\pi^2$,
\begin{equation}
m_{\rm reduced}=16\pi^2 m_{\rm phys}
\end{equation}

The first three Tables~\ref{compare115a}, \ref{comparescalar115a} and \ref{comparespectrum115a} summarize our results for the mass spectrum
at the high scale for meson-deformed 1-1-5 model specified in Table~\ref{tabvev115a}.

\begin{table}[!h]
\begin{center}
\begin{tabular}{|c|c|c|c|c|}
\hline Contribution (in units of $\mu_X$) &  $\rho$, $\tilde{\rho}$, $Z$, $\tilde{Z}$ &
$X$ & $\tilde{M}$ $M$ & total \tabularnewline \hline \hline
Tree-level constrained & $8.22\times 10^{-5}$ & $0$ & $0$
&$8.22\times 10^{-5}$ \tabularnewline Unconstrained (tree scalar
mass matrix)& $5.34\times10^{-3}$ & $0$ & $0$&
$5.34\times10^{-3}$\tabularnewline Unconstrained (mass matrix with
CW)& $2.81\times10^{-3}$ & $4.49\times10^{-3}$  & $8.3\times10^{-5}$
& $7.38\times10^{-3}$ \tabularnewline
\hline
\end{tabular}\par
\end{center}
\caption{\it Contributions to the reduced
gaugino mass $m_{1/2}$ for the meson-deformed 1-1-5 model of Table~\ref{tabvev115a}.}\label{compare115a}
\end{table}

\begin{table}[!h]
\begin{center}
\begin{tabular}{|c|c|}
\hline Contribution (in units of $\mu_X$) &  $\rho$, $\tilde{\rho}$, $Z$, $\tilde{Z}$
\tabularnewline \hline \hline Tree-level constrained &   $0.48$
\tabularnewline Unconstrained (tree scalar mass matrix)& $0.48$
\tabularnewline Unconstrained (mass matrix with CW)& not consistent
\tabularnewline
\hline
\end{tabular}\par
\end{center}
\caption{\it Contributions to the reduced
sfermion masses $m_0$ (only $\rho$, $\tilde{\rho}$, $Z$, $\tilde{Z}$ contribution) for the meson-deformed 1-1-5 model of Table~\ref{tabvev115a}. The third line in the table indicates that the use of the full CW
corrected masses is inappropriate in this case (see text).}\label{comparescalar115a}
\end{table}

\begin{table}[!h]
\begin{center}
\begin{tabular}{|c|c|}
\hline
Particle &  ${\rm Reduced\,Mass}/\mu_{X}$  \tabularnewline
\hline
\hline
sfermions & 0.48 \tabularnewline
gauginos & $7.4\times10^{-3}$\tabularnewline
$\chi_{f}$ &  $0.13$ \tabularnewline
$\chi_{s}$ &  $1.33$, $2.35$ \tabularnewline
$M_{f}$, $\tilde{M}_{f}$ & $0.42$  \tabularnewline
$M_{s}$, $\tilde{M}_{s}$ &  $9.58$, $9.73$ \tabularnewline
\hline
\end{tabular}\par
\end{center}
\caption{\it Reduced masses for the various particles charged under the SM gauge group for the meson-deformed 1-1-5 model of Table~\ref{tabvev115a}, with
$M_{\rm SUSY}/\mu_X=2.7$.}\label{comparespectrum115a}
\end{table}


The following four Tables~\ref{tabvev115b}, \ref{compare115b}, \ref{comparescalar115b} and \ref{comparespectrum115b} give results for the same 1-1-5 model but
with a different choice of parameters. Comparing the last lines in Table~\ref{compare115a} and Table~\ref{compare115b} we see that the contribution from
the $X$ messengers can be of the same order but the relative sizes of the different contributions can vary quite significantly.

In total both models give rather similar predictions. with scalars being two orders of magnitude heavier than the gauginos.
This is a ``slightly'' split-SUSY scenario which is expected in all of our direct mediation ISS-SSM models.

In addition, as can be seen from Tables~\ref{comparespectrum115a}, \ref{comparespectrum115b}, some of the messengers which are charged
under the Standard Model gauge group are relatively light with masses somewhere in between the scalars and the gauginos.

\begin{table}[!h]
\begin{center}
\begin{tabular}{|c|c|c|c|c|c|c|c|}
\hline
Vev &  $\kappa/\mu_{X}=\xi/\mu_{X}$ &  $\eta/\mu_{X}$& $p/\mu_{X}$ & $\chi/\mu_{X}$& $A/\mu_{X}$ & $B/\mu_{X}$ & $C/\mu_{X}$\tabularnewline
\hline
\hline
Tree-level constrained & $4.7610$ & $0$ & $-0.0881$ & $-2.5014$ & $17.1430$ & $13.6110$ & $215.92$ \tabularnewline
Unconstrained & $4.7603$ & $0.0017$ & $-0.0783$ & $-2.5127$  & $17.1978$ & $13.5634$ & $217.38$ \tabularnewline
\hline
\end{tabular}\par
\end{center}
\caption{\it Stabilized vevs for a meson model with $N_f=7$, $N_c=6$, $h=1$, $m_1/\mu_X=0.05$, $m_2/\mu_X=0.01$, $\mu_Y/\mu_X=5$, $\mu_P/\mu_X=3$ and
$\lambda=0.01$.}\label{tabvev115b}
\end{table}

\begin{table}[!h]
\begin{center}
\begin{tabular}{|c|c|c|c|c|}
\hline Contribution (in units of $\mu_X$) &  $\rho$, $\tilde{\rho}$, $Z$, $\tilde{Z}$ &
$X$ & $\tilde{M}$ $M$ & total \tabularnewline \hline \hline
Tree-level constrained & $5.91\times 10^{-5}$ & $0$ & $0$ &$5.91\times 10^{-5}$ \tabularnewline
Unconstrained (tree scalar mass matrix)& $3.45\times10^{-3}$ & $0$ & $0$& $3.45\times10^{-3}$\tabularnewline
Unconstrained (mass matrix with CW)& $1.78\times10^{-3}$ & $7.06\times10^{-4}$  & $1.34\times10^{-5}$ & $2.50\times10^{-3}$ \tabularnewline
\hline
\end{tabular}\par
\end{center}
\caption{\it Contributions to the reduced gaugino  mass $m_{1/2}$ for the meson-deformed 1-1-5 model of Table~\ref{tabvev115b}.}\label{compare115b}
\end{table}

\begin{table}[!h]
\begin{center}
\begin{tabular}{|c|c|}
\hline Contribution (in units of $\mu_X$) &  $\rho$, $\tilde{\rho}$, $Z$, $\tilde{Z}$
\tabularnewline \hline \hline Tree-level constrained &   $0.53$
\tabularnewline Unconstrained (tree scalar mass matrix)& $0.54$
\tabularnewline Unconstrained (mass matrix with CW)& not consistent
\tabularnewline
\hline
\end{tabular}\par
\end{center}
\caption{\it Contributions to the reduced sfermion masses $m_0$ (only $\rho$, $\tilde{\rho}$, $Z$, $\tilde{Z}$ contribution)
for the meson-deformed 1-1-5 model of Table~\ref{tabvev115b}.}\label{comparescalar115b}
\end{table}

\begin{table}[!h]
\begin{center}
\begin{tabular}{|c|c|}
\hline
Particle &  ${\rm Reduced\,Mass}/\mu_{X}$  \tabularnewline
\hline
\hline
sfermion & 0.54 \tabularnewline
gauginos & $2.5\times10^{-3}$\tabularnewline
$\chi_{f}$ &  $8.83\times10^{-2}$ \tabularnewline
$\chi_{s}$ &  $2.39$, $2.71$ \tabularnewline
$M_{f}$, $\tilde{M}_{f}$ & $0.24$  \tabularnewline
$M_{s}$, $\tilde{M}_{s}$ &  $10.20$, $10.16$ \tabularnewline
\hline
\end{tabular}\par
\end{center}
\caption{\it Reduced masses for the various particles charged under the SM gauge group for
the meson-deformed 1-1-5 model of Table~\ref{tabvev115b}, with $M_{\rm SUSY}/\mu_X=2.7$.}\label{comparespectrum115b}
\end{table}


The remaining six tables in this subsection give an example for a 2-2-3 model -- a model with a non-trivial magnetic group.
This model has very similar features with the only exception being that the reduced gaugino and sfermion masses differ for the different gauge groups.
This shows that one can achieve a deviation from the simple scaling of the full physical masses with the gauge
couplings Eqs.~\eqref{gaugino1} and \eqref{sferm1} because $m_{1/2}$ and $m_{0}$ now actually depend on the index $A$ specifying the gauge group.

\begin{table}[!h]
\begin{center}
\begin{tabular}{|c|c|c|c|c|c|c|c|}
\hline
Vev &  $\kappa/\mu_{X}=\xi/\mu_{X}$ &  $\eta/\mu_{X}$& $p/\mu_{X}$ & $\chi/\mu_{X}$& $A/\mu_{X}$ & $B/\mu_{X}$ & $C/\mu_{X}$\tabularnewline
\hline
\hline
Tree-level constrained & $4.5607$ & $0$ & $-1.3999$ & $-4.3327$ & $33.2544$ & $12.6299$ & $105.071$ \tabularnewline
Unconstrained & $4.5613$ & $0.0021$ & $-1.3433$ & $-4.2233$  & $33.5579$ & $12.4704$ & $103.038$ \tabularnewline
\hline
\end{tabular}\par
\end{center}
\caption{\it Stabilized vevs for a meson model with $N_f=7$, $N_c=5$,
$h=1$, $m_1/\mu_X=0.03$, $m_2/\mu_X=0.05$, $\mu_Y/\mu_X=5$, $\mu_P/\mu_X=3$ and
$\lambda=0.01$.}\label{tabvev223}
\end{table}

\begin{table}[!h]
\begin{center}
\begin{tabular}{|c|c|c|c|c|}
\hline Contribution (in units of $\mu_X$) & $\sigma$, $\tilde{\sigma}$, $N$, $\tilde{N}$ & $P$ & $\tilde{M}$ $M$ & total \tabularnewline \hline \hline
Tree-level constrained & $-4.5\times 10^{-3}$ & $0$ & $0$ &$-4.5\times 10^{-3}$ \tabularnewline
Unconstrained (tree scalar mass matrix)& $4.1\times10^{-3}$ & $0$ & $0$& $-4.4\times10^{-3}$\tabularnewline
Unconstrained (mass matrix with CW)& $-4.52\times10^{-2}$ & -$4.17\times10^{-4}$  & $-3.6\times10^{-4}$ & $-4.60\times10^{-2}$ \tabularnewline
\hline
\end{tabular}\par
\end{center}
\caption{\it Contributions to the reduced mass $m^{(2)}_{1/2}$
of the SU(2) gaugino for the meson-deformed 2-2-3 model of
Table~\ref{tabvev223}.}\label{compare223SU2}
\end{table}

\begin{table}[!h]
\begin{center}
\begin{tabular}{|c|c|c|c|c|}
\hline Contribution (in units of $\mu_X$) &  $\rho$, $\tilde{\rho}$, $Z$, $\tilde{Z}$ & $\chi$ & $\tilde{M}$ $M$ & total \tabularnewline \hline \hline
Tree-level constrained & $2.8\times 10^{-3}$ & $0$ & $0$ &$2.8\times 10^{-3}$ \tabularnewline
Unconstrained (tree scalar mass matrix)& $1.1\times10^{-2}$ & $0$ & $0$& $1.1\times10^{-2}$\tabularnewline
Unconstrained (mass matrix with CW)& $1.05\times10^{-2}$ & $1.05\times10^{-2}$  & $-2.4\times10^{-4}$ & $2.1\times10^{-2}$ \tabularnewline
\hline
\end{tabular}\par
\end{center}
\caption{\it Contributions to the reduced gluino mass $m^{(3)}_{1/2}$
for the meson-deformed 2-2-3 model of Table~\ref{tabvev223}.}\label{compare223SU3}
\end{table}

\begin{table}[!h]
\begin{center}
\begin{tabular}{|c|c|}
\hline
Contribution (in units of $\mu_X$) &  $\sigma$, $\tilde{\sigma}$, $N$, $\tilde{N}$  \tabularnewline
\hline
\hline
Tree-level constrained &   $2.93$ \tabularnewline
Unconstrained (tree scalar mass matrix)& $2.94$ \tabularnewline
Unconstrained (mass matrix with CW)& not consistent \tabularnewline
\hline
\end{tabular}\par
\end{center}
\caption{\it Contributions to the reduced masses $m_0^{(2)}$ of
the SU(2) sfermions (only $\rho$, $\tilde{\rho}$, $Z$, $\tilde{Z}$ contribution) for the meson-deformed 2-2-3 model of Table~\ref{tabvev223}.}\label{comparescalars223SU2}
\end{table}

\begin{table}[!h]
\begin{center}
\begin{tabular}{|c|c|}
\hline
Contribution (in units of $\mu_X$) &  $\rho$, $\tilde{\rho}$, $Z$, $\tilde{Z}$  \tabularnewline
\hline
\hline
Tree-level constrained &   $1.74$ \tabularnewline
Unconstrained (tree scalar mass matrix)& $1.74$ \tabularnewline
Unconstrained (mass matrix with CW)& not consistent \tabularnewline
\hline
\end{tabular}\par
\end{center}
\caption{\it Contributions to the SU(3) sfermion masses $m_0^{(3)}$
(only $\sigma$, $\tilde{\sigma}$, $N$, $\tilde{N}$ contribution) for the meson-deformed 2-2-3 model of Table~\ref{tabvev223}.}\label{comparescalars223SU3}
\end{table}

\begin{table}[!h]
\begin{center}
\begin{tabular}{|c|c|}
\hline
Particle &  ${\rm Reduced\,Mass}/\mu_{X}$  \tabularnewline
\hline
\hline
sfermions SU(2) &  2.95 \tabularnewline
sfermions SU(3) & 1.74 \tabularnewline
gauginos SU(2) & $4.6\times10^{-2}$ \tabularnewline
gauginos SU(3) & $2.1\times10^{-2}$\tabularnewline
$\chi_{f}$ &  $0.41$ \tabularnewline
$\chi_{s}$ &  $14.46$, $15.06$ \tabularnewline
$P_{f}$ &  $0.62$ \tabularnewline
$P_{s}$ &  $5.40$, $8.56$ \tabularnewline
$M_{f}$, $\tilde{M}_{f}$ & $0.47$  \tabularnewline
$M_{s}$, $\tilde{M}_{s}$ &  $11.79$, $11.56$ \tabularnewline
\hline
\end{tabular}\par
\end{center}
\caption{\it Reduced masses for the various particles charged under the SM gauge group for the meson-deformed 2-2-3 model of Table~\ref{tabvev223}
with $M_{\rm SUSY}/\mu_X=2.96$.}\label{comparespectrum223}
\end{table}
We have generated the soft SUSY breaking terms of the SSM at the high (messenger) scale.
In order to determine the
mass spectrum at the electroweak scale the soft SUSY breaking parameters given in the tables should be renormalization group evolved. But we expect that
the overall pattern remains the same.

In summary, we see that all our direct models have the following features: 1) A heavy scalar spectrum; 2) The pseudo-Goldstone direct messengers are relatively light
and the effective low energy theory is always extended away from the MSSM; 3) We can have deviations from the standard gaugino/sfermion mass pattern dictated by the
Standard Model gauge couplings.


\section{The baryon-deformed ISS theory and its mediation patterns} \label{sec:baryon-d}

In this Section we revisit models with the hidden sector given by baryon-deformed ISS theory introduced in \cite{Abel:2007jx,Abel:2007nr}.
These models form extensions/deformations of the ISS which are complimentary to the meson deformations discussed above.
We will extend the analysis to include the effects of the $X$ and $M$ messengers.

\subsection{The baryon-deformed model} \label{sec:baryon-d-m}

We start with an ISS model with $N_{c}=5$ colours and
$N_{f}=7$ flavours, which has a magnetic dual description as an $SU(2)$ theory,
also with $N_{f}=7$ flavours and
following \cite{Abel:2007jx,Abel:2007nr} we deform this theory by the
addition of a baryonic operator.
The resulting superpotential is given by
\begin{equation}
W=\Phi_{ij}\varphi_{i}.\tilde{\varphi_{j}}-\mu_{ij}^{2}\Phi_{ji}+m\varepsilon_{ab}
\varepsilon_{rs}\varphi_{r}^{a}\varphi_{s}^{b}\,\,
\label{Wbardef}
\end{equation}
where $i,j=1...7$ are flavour indices, $r,s=1,2$ run over the first
two flavours only, and $a,b$ are $SU(2)$ indices.
This is the superpotential of ISS with the
exception of the last term which is a baryon of the magnetic $SU(2)$ gauge group. Note that
the 1,2 flavour indices and the 3...7 indices have a different status
and the full flavour symmetry $SU(7)_f$ is broken explicitly to $SU(2)_{f}\times SU(5)_{f}$.
As before, the {\it direct} gauge mediation is implemented by
gauging the $SU(5)_{f}$ factor and identifying it
with the parent $SU(5)$ gauge group of the Standard Model.
The matter field decomposition under the magnetic $SU(2)_{gauge} \times SU(5)_{f}\times SU(2)_{f}$ and their $U(1)_R$ charges are given
in Table~\ref{fieldstableM} with $R=1$.

Using the notation established in the previous sections for the meson model the baryon-deformed model defined by Eq.~\eqref{Wbardef}
is a 2-5 model. It is straightforward to consider alternatives such as a 1-5 model where the magnetic gauge group is empty and the baryon deformation is a linear
operator,
\begin{equation}
W_{1-5}=\Phi_{ij}\varphi_{i}.\tilde{\varphi_{j}}-\mu_{ij}^{2}\Phi_{ji}+k \varphi_{1},
\label{Wbardefsimple}
\end{equation}
or, for example, a 2-2-3 model as before.
In all of those models Landau poles inherent in the direct mediation can be avoided by using the deflected unification mechanism of \cite{Abel:2008tx}.
This works most effectively in the 1-5 model due to its minimal matter content. The discussion of these models is virtually identical to that which
we will now present for the 2-5 model.

At the Lagrangian level this baryon-deformed model respects $R$-symmetry.
Thanks to the baryon deformation, the structure
of $R$-charges allows for spontaneous $R$ symmetry breaking
and it was shown in \cite{Abel:2007jx} that this does indeed happen.
We also stress that our baryon deformation is the leading order deformation of
the ISS model that is allowed by $R$-symmetry of the full theory imposed at the Lagrangian level.
As explained in \cite{Abel:2007nr} this is a self-consistent approach.
For example, terms quadratic in the
meson $\Phi$ that could arise from lower dimensional irrelevant operators
in the electric theory are forbidden by $R$-symmetry.
Thus, our deformation is described by a \emph{generic}
superpotential and \eqref{Wbardef} gives its leading-order terms.

Using the $SU(2)_{f}\times SU(5)_{f}$ symmetry, the matrix
$\mu_{ij}^{2}$ can be brought to the form \eqref{25mu2}.
The baryon operator can
be identified with a corresponding operator in the electric theory.
Indeed the mapping from baryons $B_{E}$ in the electric theory to
baryons $B_{M}$ of the magnetic theory, is $B_{M}\Lambda_{\sst ISS}^{-N}\leftrightarrow
B_{E}\Lambda_{\sst ISS}^{-N_{c}}$
(we neglect factors of order one). Thus one expects
\begin{equation}
m\sim
M_{Pl}\left(\frac{\Lambda_{\sst ISS}}{M_{Pl}}\right)^{N_{f}-2N}=\frac{\Lambda_{\sst ISS}^{3}}{M_{Pl}^
{2}},
\label{mbardef}
\end{equation}
where $M_{Pl}$ represents the scale of new physics in the electric
theory at which the irrelevant operator $B_{M}$ is generated.

The $F$-term contribution
to the potential at tree-level is
\begin{eqnarray}
V & = & \sum_{ar}|Y_{rs}\tilde{\phi}_{s}^{a}+Z_{r\hat{i}}\tilde{\rho}_{\hat{i}}^{a}
+2m\varepsilon_{ab}\varepsilon_{rs}\phi_{s}^{b}|^{2}
\\\nonumber
&&\!\!\!\!\!\!+\sum_{a\hat{i}}|\tilde{Z}_{\hat{i}r}\tilde{\phi}_{r}^{a}+X_{\hat{i}\hat{j}}\tilde{\rho}_{\hat{j}}^{a}|^{2}+
\sum_{as}|\phi_{r}^{a}Y_{rs}+\rho_{\hat{i}}^{a}\tilde{Z}_{\hat{i}s}|^{2}
+\sum_{a\hat{j}}|\phi_{r}^{a}Z_{r\hat{j}}+\rho_{\hat{i}}^{a}X_{\hat{i}\hat{j}}|^{2}
\\\nonumber
&&\!\!\!\!\!\!+
 \sum_{rs}|(\phi_{r}.\tilde{\phi_{s}}-\mu_{Y}^{2}\delta_{rs})|^{2}+\sum_{r\hat{i}}|\phi_{r}.\tilde{\rho}_{\hat{i}}|^{2}
+\sum_{r\hat{i}}|\rho_{\hat{i}}.\tilde{\phi}_{s}|^{2}+\sum_{\hat{i}\hat{j}}|(\rho_{\hat{i}}.\tilde{\rho}_{\hat{j}}
-\mu_{X}^{2}\delta_{\hat{i}\hat{j}})|^{2}
 \end{eqnarray}
where $a,\, b$ are $SU(2)_{mg}$ indices. The flavor indices $r,\,s$ and $\hat{i},\hat{\, j}$
 correspond to the $SU(2)_{f}$ and $SU(5)_{f}$, respectively.
It is straightforward to see that the rank condition
works as in ISS; that is the minimum for a given value of $X,Y,Z$
and $\tilde{Z}$ is along $\rho=\tilde{\rho}=0$ and
\be
\label{tree1}
\vev{\phi}  =  \frac{\mu_{Y}^{2}}{\xi}\,\mathbf{I}_{2} \ , \qquad
\vev{\tilde{\phi}}  =  \xi\,\mathbf{I}_{2},
\ee
where $\xi$ parameterizes a runaway direction that will
be stabilized by the Coleman-Weinberg potential Eq.~\eqref{CW}.
This then gives
$Z=\tilde{Z}=0$. In addition $Y$ becomes
diagonal and real (assuming $m$ is real).
Defining $\vev{Y_{rs}}=\eta\,\mathbf{I}_{2}$, the full potential is
\begin{equation}
\label{potential}
V=2\left|\eta\,\xi+2m\frac{\mu_{Y}^{2}}{\xi}\right|^{2}+2\left|\eta\frac{\mu_{Y}^{2}}{\xi}\right|^{2}+5\mu_{X}^{4}.
\end{equation}
Using $R$ symmetry we can choose $\xi$ to be real\footnote{The phase of $\xi$ corresponds to the $R$-axion.}.
Minimizing in $\eta$ we find
\begin{equation}
\label{tree2}
\eta=-2m\left(\frac{\xi^{2}}{\mu_{Y}^{2}}+\frac{\mu_{Y}^{2}}{\xi^{2}}\right)^{-1}.
\end{equation}
Substituting $\eta(\xi)$ into Eq.~\eqref{potential} we see that $\xi\rightarrow\infty$ is a runaway direction along which
\begin{equation}
V(\xi)=8m^{2}\mu_{Y}^{2}\left(\frac{\xi^{6}}{\mu_{Y}^{6}}+\frac{\xi^{2}}{\mu_{Y}^{2}}\right)^{-1}+5\mu_{X}^{4}.
\end{equation}
Since in the limit $\xi\rightarrow\infty$, the scalar potential $V$ is non-zero,  we have a
runaway to broken supersymmetry, hence the Coleman-Weinberg
potential again lifts and stabilizes this direction, which is indeed the case \cite{Abel:2007jx}.
As in Eqs.~\eqref{yvevs0} we parameterise the pseudo-Goldstone and runaway VEVs by
\bea
\label{phivevs}
\vev{\tilde\phi} &=& \xi\,\mathbf{I}_{2}\quad\quad\quad\quad\,\,\vev{\phi}=\kappa\,\mathbf{I}_{2}\\
\label{yvevs}
\vev{Y} &=& \eta\,\mathbf{I}_{2}\quad\quad\quad\quad\vev{X}=\chi\,\mathbf{I}_{5}.
\eea
Stabilized VEVs for a 2-5 and a 1-5 model are shown in Tables~\ref{tabvev25} and \ref{tabvev15}, respectively.
Constrained VEVs in these tables arise from using the tree-level equations of motion Eqs.~\eqref{tree1} and \eqref{tree2}.
Again, the difference between constrained and unconstrained VEVs is rather small but the general discussion of subsection \ref{sec:meson-dir}
indicates that this difference has crucial effects on the generation of gaugino masses in direct mediation.

Explicit mediation has been studied in \cite{Abel:2007nr} and leads to the usual standard GMSB pattern (as also discussed for the meson-deformed
model in subsection \ref{sec:meson-ex}).

\begin{table}[!h]
\begin{center}
\begin{tabular}{|c|c|c|c|c|}
\hline
Vev &  $\kappa/\mu_{X}$ & $\xi/\mu_{X}$ &  $\eta/\mu_{X}$& $\chi/\mu_{X}$  \tabularnewline
\hline
\hline
Tree-level constrained & $1.1005$ & $8.1781$ & $-0.0793$ & $-0.3493$  \tabularnewline
Unconstrained & $1.1004$ & $8.1766$  & $-0.0792$  & $-0.3470$  \tabularnewline
\hline
\end{tabular}\par
\end{center}
\caption{\it Stabilized VEVs for a 2-5 baryon-deformed model with $N_f=7$, $N_c=5$, $h=1$, $m/\mu_{X}=0.3$ and $\mu_Y/\mu_{X}=3$.}\label{tabvev25}
\end{table}

\begin{table}[!h]
\begin{center}
\begin{tabular}{|c|c|c|c|c|}
\hline
Vev &  $\kappa/\mu_{X}$ & $\xi/\mu_{X}$ &  $\eta/\mu_{X}$& $\chi/\mu_{X}$  \tabularnewline
\hline
\hline
Tree-level constrained & $1.76214$ & $5.1074$ & $-0.05248$ & $-0.20720$  \tabularnewline
Unconstrained & $1.7620$ & $5.1067$  & $-0.05227$  & $-0.2037$  \tabularnewline
\hline
\end{tabular}\par
\end{center}
\caption{\it Stabilized VEVs for a 1-5 baryon-deformed model with $N_f=6$, $N_c=5$, $h=1$, $k/\mu^{2}_{X}=0.3$ and $\mu_Y/\mu_{X}=3$.}\label{tabvev15}
\end{table}

\subsection{Summary of signatures in the directly mediated baryon-deformed model} \label{sec:baryon-dir}

The basic equations for calculating gaugino and scalar masses are the same as in subsection~\ref{sec:meson-dir}.
Only the VEV configurations and the structure of the messenger mass matrices know about the difference in the deformation.

Our results for the soft SUSY breaking parameters at the messenger scale are presented below following the same structure as before.
The first three tables correspond to the 2-5 model given in Table~\ref{tabvev25}. The next three correspond to the 1-5 model specified in Table~\ref{tabvev15}.

Evidently, the dominant contribution to the gaugino mass comes from unconstraining the VEVs and putting in the full one-loop mass matrices.
Overall this leads again to models with heavy scalars and, in distinction to our earlier paper~\cite{Abel:2007nr} (where the
constrained VEVs were used), we do not need to fine tune
the different $\mu^2$ parameters to achieve a moderately split spectrum. It is remarkable that in all of the directly mediated ISS models gaugino masses are
this sensitive to quantum corrections (due to the inevitable cancellation at tree-level).

\begin{table}[!h]
\begin{center}
\begin{tabular}{|c|c|c|c|}
\hline Contribution &  $\rho$, $\tilde{\rho}$, $Z$, $\tilde{Z}$ & $\chi$  & total \tabularnewline \hline \hline
Tree-level constrained & $4.17\times 10^{-5}$ & $0$  &$4.17\times 10^{-5}$ \tabularnewline
Unconstrained (tree scalar mass matrix)& $1.74\times10^{-3}$ & $0$ & $1.74\times10^{-3}$\tabularnewline
Unconstrained (mass matrix with CW)& $-1.57\times10^{-3}$ & $9.61\times10^{-7}$  &  $-1.57\times10^{-3}$ \tabularnewline
\hline
\end{tabular}\par
\end{center}
\caption{\it Contributions to the reduced gaugino mass for the baryon-deformed 2-5 model of Table~\ref{tabvev25}.}\label{compare25}
\end{table}

\begin{table}[!h]
\begin{center}
\begin{tabular}{|c|c|}
\hline
Contribution &  $\rho$, $\tilde{\rho}$, $Z$, $\tilde{Z}$  \tabularnewline
\hline
\hline
Tree-level constrained &   $0.70$ \tabularnewline
Unconstrained (tree scalar mass matrix)& $0.70$ \tabularnewline
Unconstrained (mass matrix with CW)& not consistent \tabularnewline
\hline
\end{tabular}\par
\end{center}
\caption{\it Contributions to the reduced sfermion masses (only $\rho$, $\tilde{\rho}$, $Z$, $\tilde{Z}$ contribution) for
the baryon-deformed 2-5 model of Table~\ref{tabvev25}.}\label{comparescalar25}
\end{table}

\begin{table}[!h]
\begin{center}
\begin{tabular}{|c|c|}
\hline
Particle &  ${\rm Mass}/\mu_{P}$  \tabularnewline
\hline
\hline
sfermion & 0.70 \tabularnewline
gauginos & $1.57\times10^{-3}$\tabularnewline
$\chi_{f}$ &  $1.92\times10^{-2}$ \tabularnewline
$\chi_{s}$ &  $2.923$, $2.925$ \tabularnewline
\hline
\end{tabular}\par
\end{center}
\caption{\it Reduced masses for the various particles charged under the SM gauge group for the baryon-deformed 2-5 model of Table~\ref{tabvev25}.}\label{comparespectrum25}
\end{table}

\begin{table}[!h]
\begin{center}
\begin{tabular}{|c|c|c|c|}
\hline Contribution &  $\rho$, $\tilde{\rho}$, $Z$, $\tilde{Z}$ & $\chi$  & total \tabularnewline \hline \hline
Tree-level constrained & $2.67\times 10^{-5}$ & $0$  &$2.67\times 10^{-5}$ \tabularnewline
Unconstrained (tree scalar mass matrix)& $7.49\times10^{-4}$ & $0$ & $7.49\times10^{-4}$\tabularnewline
Unconstrained (mass matrix with CW)& $-5.97\times10^{-4}$ & $3.60\times10^{-7}$  &  $-5.96\times10^{-4}$ \tabularnewline
\hline
\end{tabular}\par
\end{center}
\caption{\it Contributions to the reduced gaugino mass for the baryon-deformed 1-5 model of Table~\ref{tabvev15}.}\label{compare15}
\end{table}

\begin{table}[!h]
\begin{center}
\begin{tabular}{|c|c|}
\hline
Contribution &  $\rho$, $\tilde{\rho}$, $Z$, $\tilde{Z}$  \tabularnewline
\hline
\hline
Tree-level constrained &   $0.61$ \tabularnewline
Unconstrained (tree scalar mass matrix)& $0.61$ \tabularnewline
Unconstrained (mass matrix with CW)& not consistent \tabularnewline
\hline
\end{tabular}\par
\end{center}
\caption{\it Contributions to the reduced sfermion masses (only $\rho$, $\tilde{\rho}$, $Z$, $\tilde{Z}$ contribution) for the baryon-deformed 1-5 model of
Table~\ref{tabvev15}.}\label{comparescalar15}
\end{table}
\begin{table}[!h]
\begin{center}
\begin{tabular}{|c|c|}
\hline
Particle &  ${\rm Mass}/\mu_{P}$  \tabularnewline
\hline
\hline
sfermions &  0.61 \tabularnewline
gauginos & $5.96\times10^{-4}$\tabularnewline
$\chi_{f}$ &  $1.1\times10^{-2}$ \tabularnewline
$\chi_{s}$ &  $2.921$, $2.919$ \tabularnewline
\hline
\end{tabular}\par
\end{center}
\caption{\it Reduced masses for the various particles charged under the SM gauge group for the baryon-deformed 1-5 model of Table~\ref{tabvev15}.}\label{comparespectrum15}
\end{table}
\newpage~
\section{Conclusions} \label{sec:concl}

We have investigated different scenarios of gauge mediation which incorporate a dynamical SUSY breaking (DSB)
sector coupled to a supersymmetric Standard Model.
The DSB sector was realized in terms of two different types of deformations of the ISS model.
These models generate all SUSY breaking parameters at the messenger scale in a calculable way from relatively simple supersymmetric Lagrangians.
In all of the models investigated we find
rather model independent signatures for the direct gauge mediation which include:
\begin{itemize}
\item{} Scalars are typically two orders of magnitude or more heavier than gauginos.
\item{} The low energy effective theory of the visible sector i.e. particles charged under the Standard Model gauge groups is necessarily extended
by light pseudo-Goldstone messenger fields.
\item{} Direct mediation models easily allow for deviations from the mass patterns dictated by the gauge couplings, familiar from standard gauge mediation.
\end{itemize}
It is also possible to implement indirect gauge mediation, by adding an explicit messenger sector.
In this case we find a rather standard pattern of gauge mediated supersymmetry breaking.

Finally we would like to briefly comment on how the usual little
hierarchy problem of the supersymmetric Standard Model manifests
itself. First of all, the non-observation of the
Higgs at LEP requires that the mass of the lightest Higgs,
$m_{h^0} > 115\, {\rm GeV}$. On the other hand, supersymmetric
models predict an upper bound so that
\be
\label{bound}
(115 \, {\rm GeV})^2 \, < \, m_{h^0}^2 \, < \, \cos^2(2\beta) m_Z^2 \, +\, {\rm rad.\, corr.}
\ ,
\ee
where the radiative corrections $\sim m_t^2 \log (m_{\tilde{t}} /m_t)$.
To fulfill this one needs a rather large stop mass, which our models deliver.
On the other hand, the conditions for electroweak symmetry breaking require that at the electroweak scale
\be
m^2_Z=-2(m^2_{H_{u}}+|\mu_{\rm MSSM}|^2)+{\mathcal{O}}(1/\tan^2(\beta)).
\ee
The scalar masses, including $m_{H_{u}}$, and their loop corrections are of the order of $m_{\tilde{t}}$ and are
(as just argued) much bigger than the electroweak scale. This requires a fine-tuning of $\mu_{\rm MSSM}$ of the order of $10^{-2}$.
In the direct mediation scenarios with a mildly split SUSY spectrum, $m_{\tilde{t}}$ is bigger than the minimal required value from
Eq.~\eqref{bound} resulting in a somewhat higher degree of fine-tuning of the order of $10^{-4}-10^{-5}$.
In this paper we are treating $\mu_{\rm MSSM}$ as a free parameter and do not attempt to solve this problem.

\section*{Acknowledgments}
VVK is supported in part by a Leverhulme Research Fellowship and LM acknowledges an FCT Postgraduate Studentship.

\bigskip

\begin{appendix}
\section{Leading order contribution to the gaugino mass}

To develop a perturbative approximation of Eqs.~\eqref{eq:master}-\eqref{intIab}
we note that when the $F$-terms
are small compared to $\mu^{2}$, we may first go to the {}``fermion-diagonal
basis'', by making a rotation on the scalars given by
\begin{equation}
Q_{0}=\left(\begin{array}{cc}
U & \mathbf{0}\\
\mathbf{0} & V\end{array}\right)
\end{equation}
where the $U$ and $V$ matrices are the fermion-diagonalisation matrices defined in \eqref{UdVdef}.
In this basis the scalar mass-squareds are
\begin{equation}
\tilde{m}_{\rm sc}^{2}=Q_{0}^{\dagger}m_{\rm sc}^{2}Q_{0}\approx
\left(\begin{array}{cc}
\hat{m}_{\rm f}^{2} & {\cal A}\\
{\cal A}^{\dagger} & \hat{m}_{\rm f}^{2}\end{array}\right)
\end{equation}
where
\begin{equation}
{\cal A}_{ij}=U_{ia}^{\dagger}W^{abc}W_{c}V_{bj}=(U^{\dagger}{\cal {F}}V)_{ij}
 \label{calA}
\end{equation}
in terms of the $F$-term matrix ${\cal F}^{ab} \equiv W^{abc}W_{c}$.
Evaluating the diagram for the gaugino mass in this basis (cf. Eqs.~\eqref{eq:master}-\eqref{intIab})
and suppressing the overall factor $2g^{2}_{A}\tr(T^{A}T^{B})$,
yields,
\begin{eqnarray}
\nonumber
&&\!\!\!\!\!\!\int \frac{d^{4}k}{(2\pi)^{4}}\sum^{4}_{k,l=1}\sum^{2}_{i,j=1}(U_{i1}^{\dagger}Q_{0,1k}+U_{i2}^{\dagger}Q_{0,2k})
\left(\frac{1}{k^{2}-\tilde{m}^{2}_{\rm sc}}\right)_{kl}
\left(\frac{\hat{m}_{f}}{k^{2}-\hat{m}^{2}_{\rm f}}\right)_{ij}(Q_{0,l3}^{\dagger}V_{1j}+Q_{0,l4}^{\dagger}V_{2j})
\\\nonumber
&\!\!\!=\!\!\!& \int \frac{d^{4}k}{(2\pi)^{4}}\sum^{2}_{i,j,k,l=1}(U_{i1}^{\dagger}U_{1k}+U_{i2}^{\dagger}U_{2k})
\left(\frac{1}{k^{2}-\tilde{m}^{2}_{\rm sc}}\right)_{k,(l+2)}
\left(\frac{\hat{m}_{f}}{k^{2}-\hat{m}^{2}_{\rm f}}\right)_{ij}(V_{l1}^{\dagger}V_{1j}+V_{l2}^{\dagger}V_{2j})
\\
&\!\!\!=\!\!\!&\int \frac{d^{4}k}{(2\pi)^{4}}\sum^{2}_{i,j,k,l=1}\delta_{ik}
\left(\frac{1}{k^{2}-\tilde{m}^{2}_{\rm sc}}\right)_{k,(l+2)}\left(\frac{\hat{m}_{f}}{k^{2}-\hat{m}^{2}_{\rm f}}\right)_{ij}\delta_{jl},
\end{eqnarray}
where, in the last step, we have made use of the unitarity of the $U$ and $V$ matrices.

The fermion propagator is already diagonal, but the boson propagator has off diagonal terms $\sim{\cal A}$. Expanding in powers
of ${\cal A}$ we have,
\begin{equation}
\label{bosonexpand}
\left(\frac{1}{k^{2}-\tilde{m}^{2}_{\rm sc}}\right)_{k,(l+2)}
=\left(\frac{1}{k^{2}-\hat{m}^{2}_{\rm f}}\mathcal{A}\frac{1}{k^{2}-\hat{m}^{2}_{\rm f}}
+\frac{1}{k^{2}-\hat{m}^{2}_{\rm f}}\mathcal{A}\frac{1}{k^{2}-\hat{m}^{2}_{\rm f}}\mathcal{A^{\dagger}}
\frac{1}{k^{2}-\hat{m}^{2}_{\rm f}}\mathcal{A}\frac{1}{k^{2}-\hat{m}^{2}_{\rm f}}+\cdots \right)_{kl}
\end{equation}
Using that $\hat{m}_{\rm f}$ is a diagonal matrix we find to lowest order in ${\cal A}$,
\begin{equation}
M_{\lambda^{A}}=2g_{A}^2\, \tr(T^{A}T^{B})\Tr({\cal A} I^{(1)}(\hat{m}_{\rm f}))
\end{equation}
where
\begin{equation}
I^{(1)}_{ij}={\rm diag}(I(\hat{m}_{ii}))
\end{equation}
and
\begin{equation}
I^{(1)}(m)=\int \frac{d^{4} k}{(2\pi)^{4}} \frac{m}{(k^{2}-m^{2})^{3}}=\frac{1}{32\pi^2}\frac{1}{m}.
\end{equation}
Using the explicit form of $I^{(1)}$ we have the leading order contribution to the gaugino masses:
\begin{equation}
\label{gauginofirst}
M_{\lambda^{A}}=\frac{g_{A}^2}{16\pi^2}\, \tr(T^{A}T^{B})\Tr({\cal A} \hat{m}^{-1}_{\rm f})
=\frac{g_{A}^2}{16\pi^2}\, \tr(T^{A}T^{B})\Tr({\cal F} m^{-1}_{\rm f})\, .
\end{equation}
This reproduces Eq.~\eqref{gaugino-one}.

\end{appendix}

\end{document}